\documentclass[a4paper,11pt]{article}
\pdfoutput=1
\usepackage{pos}
\usepackage{setspace}

\usepackage{tikz-cd}
\tikzcdset{arrow style=math font}

\newcommand{\bep}{\begin{picture}}
\newcommand{\eep}{\end{picture}}

\newcounter{YoungHeight}\newcounter{YoungWidth}

\newcounter{Mul1}\newcounter{Mul2}\newcounter{Mul3}\newcounter{Mul4}
\newcounter{A0}\newcounter{A1}\newcounter{A2}
\newcounter{B3}
\newcounter{C3}\newcounter{C4}
\newcounter{D1}\newcounter{D2}\newcounter{D3}
\newcounter{T0}\newcounter{T1}

\newlength{\txtHShift}

\newlength{\txtWidth}

\usepackage{skak}
\usepackage{mathtools}
\usepackage{graphicx}
\newcommand{\intprod}{\mathbin{\raisebox{\depth}{\scalebox{1}[-1]{$\lnot$}}}}

\newcommand{\Kcurl}{\mathscr{K}}
\newcommand{\TA}{\mathtt{A}}
\newcommand{\hatphi}{\hat\phi}
\newcommand{\TB}{\mathtt{B}}
\newcommand{\ty}{\text{y}}
\newcommand{\tx}{\text{x}}
\newcommand{\diag}{\text{diag}}
\newcommand{\Mcurl}{\mathscr M}

\newcommand{\ellbold}{\boldsymbol \ell}
\newcommand{\Dc}{\mathrm{D}}
\newcommand{\Jbold}{\boldsymbol J}
\newcommand{\Tr}{\text{Tr}}

\newcommand{\sa}{\mathsf{a}}

\newcommand{\cI}{\mathcal{I}}
\newcommand{\cJ}{\mathcal{J}}

\newcommand{\hatSigma}{\hat \Sigma}

 \def\one{\mbox{1 \kern-.59em {\rm l}}}

\newcommand{\msp}{\mathfrak{sp}}
\newcommand{\eps}{\epsilon}

\newcommand{\sF}{\mathsf{F}}

\newcommand{\msu}{\mathfrak{su}}

\newcommand{\C}{\mathbb{C}}

\newcommand{\PT}{\mathbb{PT}}

\newcommand{\PS}{\mathbb{PS}}
\newcommand{\R}{\mathbb{R}}
\renewcommand{\P}{\mathbb{P}}

\newcommand{\N}{\mathbb{N}}

\newcommand{\Z}{\mathbb{Z}}

\newcommand{\A}{\mathbb{A}}

\newcommand{\e}{\mathrm{e}}

\newcommand{\pl}{\partial}

\newcommand{\cE}{\mathcal{E}}

\newcommand{\cH}{\mathcal{H}}
\newcommand{\cM}{\mathcal{M}}
\newcommand{\cN}{\mathcal{N}}
\newcommand{\cO}{\mathcal{O}}
\newcommand{\cA}{\mathcal{A}}

\newcommand{\cL}{\mathcal{L}}

\newcommand{\cB}{\mathcal{B}}
\newcommand{\cC}{\mathcal{C}}

\newcommand{\cZ}{\mathcal{Z}}

\newcommand{\hs}{\mathfrak{hs}}
\newcommand{\tK}{\mathtt{K}}

\newcommand{\PTc}{\mathcal{PT}}

\newcommand{\mso}{\mathfrak{so}}

\title{Twistor approach to higher-spin theories and matrix model}
\DeclareUnicodeCharacter{2212}{\ensuremath{-}}
\author*{Tung Tran}

\affiliation{Service de Physique de l’Univers, Champs et Gravitation \\
    Universit\'e de Mons, 20 place du Parc, 7000 Mons, Belgium}

\emailAdd{vuongtung.tran@umons.ac.be}

\abstract{We discuss recent endeavours in connecting twistor theory to higher-spin theories and the IKKT-matrix model. Starting with a brief review on higher-spin algebra $\hs$ in four-dimensional target space, we elucidate how higher-spin symmetry can be encoded in $\hs$-valued sections/holomorphic differential forms on (non-commutative) twistor space. This provides an efficient way to construct local higher-spin theories in spacetime from some actions on (non-commutative) twistor space. Remarkably, some higher-spin theories obtained within the framework of twistor theory can have non-trivial scattering amplitudes in flat space.}

\FullConference{Corfu Summer Institute 2022 "School and Workshops on Elementary Particle Physics and Gravity"\\
September 18 - September 25, 2022 \\
Corfu, Greece}

\begin{document}

\maketitle

\section{Introduction}
Not so long after the discovery of twistor theory \cite{Penrose:1967wn}, people realized that there is a profound relation between \cite{mason1996integrability}
\begin{center}
    \textbf{Integrable systems in $4d$ spacetime} \quad $\longleftrightarrow$ \quad \textbf{Holomorphic structures on twistor space}
\end{center}

For instance, self-dual Yang-Mills (SDYM) \cite{Chalmers:1996rq} and self-dual gravity (SDGRA) \cite{Siegel:1992wd,Krasnov:2016emc,Krasnov:2021cva} can be formulated as BF and Poisson-BF actions on twistor space \cite{Bittleston:2022nfr}, respectively.\footnote{We refer the readers to e.g. \cite{Jiang:2008xw,Adamo:2017qyl,Atiyah:2017erd} for a review on twistor theory.} Despite these successes, there were not many activities in finding interacting higher-spin theories from twistor community compared to the developments in higher-spin community initiated by Fradkin-Vasiliev \cite{Fradkin:1986ka,Fradkin:1987ks}, Vasiliev \cite{Vasiliev:1990en,Vasiliev:2003ev}, Bengtsson-Bengtsson-Brink \cite{Bengtsson:1983pd,Bengtsson:1986kh} and Metsaev \cite{Metsaev:1991mt,Metsaev:1991nb}. The main technical problem is encapsulated in encoding higher-spin symmetry into some geometrical datas on the twistor space with the correct projective scaling so that we can obtain local higher-spin interactions in spacetime. It is noteworthy that free equations of motion for massless higher-spin fields have been known long time ago in twistor theory
\cite{Penrose:1967wn,Penrose:1976js} even before Fronsdal \cite{Fronsdal:1978vb} and Fang \cite{Fang:1978wz} wrote down their Lagrangians for free massless higher-spin fields.

Very roughly, if  $J_{a(s)}$ is a conserved higher-spin rank-$s$ tensor associated with the higher-spin current $\Jbold^s$ and $t^{a(s-1)}$ is a rank-$(s-1)$ conformal Killing tensor, then the corresponding higher-spin charge $Q^s$ can be defined as\footnote{We invite the readers to the report \cite{Bekaert:2022poo} and the lecture notes \cite{Didenko:2014dwa,Ponomarev:2022atv} for a review on higher-spin theories.}
\begin{align}
    Q^{s}(t)=\int d^{d-1}x \, \Jbold_0^{s}\,,\qquad \text{where}\qquad \Jbold^{s}_m(t)=J_{ma(s-1)}t^{a(s-1)}\,,
\end{align}
Since CFT axioms require, for instance
\begin{align}
    [Q_2,Q_s]=Q_s+...\,,\qquad \qquad [Q_s,Q_s]=Q_2+...\,,
\end{align}
where we have other higher-spin charges in the ellipsis, unless there are higher-spin charges of all spins (at least even) \cite{Maldacena:2011jn,Maldacena:2012sf,Boulanger:2013zza,Alba:2013yda,Alba:2015upa}, the Ward/Jacobi identities will be violated. The tower of infinitely many conserved charges $Q^s$ form an associative higher-spin algebra which we will denote as $\hs$ \cite{Fradkin:1986ka,Eastwood:2002su}. Note that $\hs$ is an extended algebra of the usual conformal algebra, which has $J_{a(2)}$ as the canonical conserved stress tensor. As a result, higher-spin symmetry requires all possible interactions between higher-spin fields in the vertices. This is the main idea behind the construction of interacting higher-spin theories.

\medskip

In 4-dimension, the cubic vertices for any given triplet of helicities $(h_1,h_2,h_3)$ can be uniquely fixed by symmetry of the little group \cite{Bengtsson:1983pd,Bengtsson:1986kh,Benincasa:2007xk}. In particular, the anti-holomorphic cubic vertices have the following form
\begin{align}\label{eq:onshell3pt}
    \bar{V}_3=\bar{C}_{h_1,h_2,h_3}[12]^{h_1+h_2-h_3}[23]^{h_2+h_3-h_1}[31]^{h_3+h_1-h_2}\,,\qquad (h_1+h_2+h_3>0)\,.
\end{align}
Note that when $h_1+h_2+h_3<0$, we simply replace the square brackets $[ij]$ by the angled brackets $\langle ij\rangle$, and remove the bar over the coupling constant $\bar{C}_{h_1,h_2,h_3}$. Chiral higher-spin gravity (HSGRA) \cite{Metsaev:1991mt,Metsaev:1991nb,Ponomarev:2016lrm} is a special class among all higher-spin theories where interactions stop at cubic order. It has the following coupling constants
\begin{align}\label{eq:couplingconstant}
    \bar{C}_{h_1,h_2,h_3}=\frac{\bar\kappa \ell_p^{h_1+h_2+h_3-1}}{\Gamma[h_1+h_2+h_3]}\,.
\end{align}
Here, $\ell_p$ has the dimension of length, and $\bar{\kappa}$ is a dimensionless parameter. It is worth to note that the coupling constants \eqref{eq:couplingconstant} have been discovered in various contexts. For instance, $\bar{C}_{h_1,h_2,h_3}$ were derived dynamically in \cite{Metsaev:1991mt,Metsaev:1991nb,Ponomarev:2016lrm,Sharapov:2022faa,Sharapov:2022awp}, while in the work of \cite{Haehnel:2016mlb,Adamo:2016ple,Tran:2022tft} $\bar{C}_{h_1,h_2,h_3}$ were understood as built-in numerical factors coming from the Taylor expansion of the Moyal-Weyl $\star$-product on twistor space. In addition, $\bar{C}_{h_1,h_2,h_3}$ were also discovered in the context of celestial amplitudes \cite{Ren:2022sws,Monteiro:2022lwm}. Since the flat space chiral HSGRA has been shown to admit a smooth deformation to its $(A)dS_4$ version \cite{Metsaev:2018xip,Skvortsov:2018uru,Sharapov:2022awp,Tran:2022tft}, it repels the common opinion that higher-spin theories can only exist in (A)dS \cite{Fradkin:1987ks}. The results of \cite{Sharapov:2022faa,Sharapov:2022awp,Tran:2022tft} have resolved the mismatch between the cubic vertices in the Fronsdal's \cite{Boulanger:2008tg} and light-cone approaches \cite{Bengtsson:1983pd,Bengtsson:1986kh,Metsaev:2005ar}. 

\medskip 

Free differential algebra approach to the construction of equations of motion for chiral higher-spin gravity (HSGRA)  \cite{Sharapov:2022faa,Sharapov:2022awp,Sharapov:2022wpz}, and the twistor construction in \cite{Haehnel:2016mlb,Adamo:2016ple,Krasnov:2021nsq,Tran:2021ukl,Steinacker:2022jjv,Tran:2022tft} are important results of the covariantization program for chiral HSGRA and its contractions from their light-cone descriptions \cite{Metsaev:1991mt,Metsaev:1991nb,Ponomarev:2016lrm,Ponomarev:2017nrr,Metsaev:2018xip,Skvortsov:2018uru}.\footnote{See also \cite{Metsaev:2019dqt,Metsaev:2019aig,Tsulaia:2022csz} for supersymmetric version of chiral HSGRA.} In this note, we want to convey a formula for: (i) constructing the dual twistor actions of various 4-dimensional local higher-spin theories, (ii) obtaining their covariant spacetime actions from (non-commutative) twistor space \cite{Haehnel:2016mlb,Adamo:2016ple,Krasnov:2021nsq,Tran:2021ukl,Steinacker:2022jjv,Tran:2022tft}. To date, most of $4d$ local HSGRAs obtained from twistor space have complex action functionals and are (quasi)-chiral type theories. Nevertheless, they are consistent theories that can avoid various No-go theorems in flat space \cite{Coleman:1967ad,Weinberg:1964ew} and AdS space \cite{Maldacena:2011jn} since some of the assumptions of the No-go theorems such as unitarity and parity invariance are violated.

Note that $4d$ (quasi-)chiral higher-spin theories tend to have simple scattering amplitudes in flat space. In fact, for quite some time there was a widespread belief that local higher-spin theories can only have trivial scattering amplitudes in flat space due to various results in \cite{Joung:2015eny,Beccaria:2016syk,Roiban:2017iqg,Ponomarev:2016cwi,Skvortsov:2018jea,Skvortsov:2020wtf,Skvortsov:2020gpn}. Depends on the audience, the triviality of higher-spin scatterings can be either intriguing or completely tedious. Therefore, it is instinctive to ask whether we can have any examples of non-trivial local higher-spin theories. As luck may have it, twistor theory allows us to expand the realm of consistent interacting higher-spin theories by perturbatively deforming away from the chiral sectors as in \cite{Haehnel:2016mlb,Adamo:2016ple,Tran:2021ukl,Steinacker:2022jjv}. In \cite{Adamo:2022lah}, it is shown that higher-spin extension of Yang-Mills theory (HS-YM) has non-trivial scattering amplitudes, which is a surprising result. In particular, the MHV amplitudes of HS-YM between two negative helicity $-s$ fields and the remaining positive helicity $+1$ fields read 
\begin{align}
    \cM(1^{+1},...,i^{-s},...,j^{-s},...,n^{+1})=\frac{\langle ij\rangle^4}{\langle 12\rangle...\langle n1\rangle}\langle ij\rangle^{2s-2}\,.
\end{align}
Observe that the above amplitude comprises a well-known Park-Taylor factor \cite{Parke:1986gb} and a part addressing two external states with negative helicity $-s$. The above amplitudes trigger a natural question of whether non-self-dual higher-spin theories can be phenomenologically significant.

\medskip 

At high energy/short distance where quantum mechanics govern physics, it is reasonable to assume that spacetime and fields should get quantized and have a unified description in one single fundamental theory. The IKKT-matrix model \cite{Ishibashi:1996xs} is a model that can offers such an opportunity to realize the above idea. It is worth to note that the IKKT is one-loop finite \cite{Steinacker:2016nsc}, and it has non-trivial connection with cosmology and black hole physics \cite{Steinacker:2019fcb}. Since the natural background of the IKKT-matrix model is a ``fuzzy'' (or quantized) twistor space $\P^3_N$ \cite{Sperling:2017dts}, it induces a higher-spin gauge theory (HS-IKKT) with a spectrum consisting of a finite number of spinning fields. In particular, if $Z^A$ is $\msp(4)$ (or $\msu(4)$) vector, and $\hat{Z}^A$ is its dual, then the space of functions on $\P^3_N$ is realized as \cite{Steinacker:2022jjv}
\begin{align}\label{eq:spectrumtruncated}
    \begin{split}
    \mathscr{C}(\P^3_N) &=End(\cH_N)=(N,0,0)_{\msu(4)}\otimes (0,0,N)_{\msu(4)} =\sum_{n=1}^{N}(n,0,n)_{\msu(4)}\,
     \\
     &=\sum_{n=0}^N f_{A(n)B(n)}Z^{A}...Z^{A}\hat{Z}^{B}...\hat{Z}^{B}\,,
     \end{split}
\end{align}
where $\cH_N=(0,0,N)=(0,0,1)^{\otimes_{\text{sym}} N}$ is an $N$-particle Fock space where $\cH_N=\hat{Z}_1^A...\hat{Z}^A_N|0\rangle$. Observe that we have a truncation of higher-spin modes on $\P^3$ \cite{Sperling:2017dts} since it is clear from \eqref{eq:spectrumtruncated} that the spectrum of the HS-IKKT on $\P^3_N$ is bounded from above.\footnote{A review on (HS)-IKKT can be found at e.g. \cite{Steinacker:2019fcb}.}  As a consequence, it is expected that the HS-IKKT can have non-trivial $S$-matrix since there is not enough symmetry to trivialize physical scattering processes; and another reason is that it is a quasi-chiral theory.
\section{Constructing chiral HSGRA from twistor space}

\paragraph{Twistor space.} If $SL(2,\C)\times SL(2,\C)\subset Sp(4,\C)$ is the local Lorentz group of a complexified conformally flat spacetime $\cM_{\C}\equiv\cM$ with cosmological constant $\Lambda$, then a null vector can be described by a pair of bosonic Weyl spinors of opposite chiralities, which live in $(\frac{1}{2},0)$ and $(0,\frac{1}{2})$
representations, i.e. $V^a=\lambda^{\alpha}\mu^{\dot\alpha}$. In what follow, we parametrize the homogeneous coordinate $Z^A$ of $\P^3$ as $Z^A=(\lambda^{\alpha},\mu^{\dot\alpha})$ where spinor indices have values $0,1$. This is in accordance with the fact that $\P^3$ can admits a spin structure since the second Stiefel-Whitney class of $\P^3$ is $w_2(T\P^3)=0$. Furthermore, since $\P^3$ is a compact symplectic manifold, there is a nature quaternionic conjugation that maps
\begin{align}
    \hat{}\,:\,Z^A=(\lambda^{\alpha},\mu^{\dot\alpha})\mapsto \hat Z^A=(\hat\lambda^{\alpha},\hat\mu^{\dot\alpha})\,,
\end{align}
such that 
\begin{align}
   \lambda^{\alpha}=(\lambda^0,\lambda^1)\mapsto \hat \lambda^{\alpha}=(-\overline{\lambda^1},\overline{\lambda^0})\,,\qquad \qquad \mu^{\dot\alpha}=(\mu^{\dot 0},\mu^{\dot 1})\mapsto \hat\mu^{\dot\alpha}=(-\overline{\mu^{\dot 1}},\overline{\mu^{\dot 0}})\,.
\end{align}
In the affine patch of $(A)dS_4$ where the metric reads
\begin{align}\label{eq:metric}
    ds^2=\frac{dx_{\mu}dx^{\mu}}{(1+\Lambda x^2)^2}=\Omega^2dx_{\mu}dx^{\mu}\,,\qquad \mu=1,2,3,4\,,
\end{align}
there is a natural object $I^{AB}$ known as the infinity twistor \cite{Penrose:1967wn} used to specify the conformal factor $\Omega$ in \eqref{eq:metric}. It has the following properties
\begin{align}
    \frac{1}{2}I^{AB}\eps_{ABCD}=I_{CD}\,,\qquad \qquad I_{AC}I^{BC}=\Lambda\delta_A{}^B\,,
\end{align}
and the following representatives
\begin{align}
    I^{AB}=\begin{pmatrix}\Lambda \eps^{\alpha\beta} & 0\\
    0&\eps^{\dot\alpha\dot\beta}\end{pmatrix}\,,\qquad \qquad I_{AB}=\begin{pmatrix}\eps_{\alpha\beta} &0\\
    0 &\Lambda \eps_{\dot\alpha\dot\beta}\end{pmatrix}\,.
\end{align}
The twistor space $\PT$ is then defined as an open subset of $\P^3$ where
\begin{align}\label{eq:PT}
    \PT=\{Z^A\in \P^3\,|\,I_{AB}Z^A\hat Z^B\neq 0\}\,.
\end{align}
Note that in the flat limit where $\Lambda\rightarrow 0$, the condition $I_{AB}Z^A\hat{Z}^B\neq 0$ reduces to the removal of the projective line $\lambda^{\alpha}=0$ (which is a point at infinity in $\cM$). 

By assuming $\mu^{\dot\alpha}=F^{\dot\alpha}(x,\lambda)$ as in \cite{kodaira1962theorem,kodaira1963stability}, we can identify the projective undotted spinor bundle $\PS\simeq \cM\times \P^1$ as the corresponding space between $\PT$ and $\cM$. This fact can be described by the double fibration:
\begin{equation}\label{eq:doublefibration}
    \begin{tikzcd}[column sep=small]
& \PS \ar["\pi_1" ',dl] \ar[dr,"\pi_2"] & \\
\PT  & & \cM
\end{tikzcd}
\end{equation}
If a twistor line over a point $x\in \cM$ is $L_x\simeq \P^1$, then the tangent space $T_x\cM$ can be identified with $H^0(L_x,N_{L_x})\simeq \C^4$ \cite{kodaira1962theorem} by virtue of a horizontal lifting. Furthermore, the normal bundle wrt. $L_x$
\begin{align}
    N_{L_x}:=T(\PT)|_{L_x}/T(L_x)\simeq \cO(1)\oplus \cO(1)\,
\end{align}
can be obtained as a consequence of Birkhoff-Grothendieck Lemma. Note that in the flat limit $\Lambda\rightarrow 0$, $\PT$ will be isomorphic to $\cO(1)\oplus \cO(1)$. 

Next, taking the advantage of the fact that $T(L_x)\cong \cO(2)$ and $T^*(L_x)\cong \cO(-2)$, where $\cO(n):=\cO(1)^{\otimes\, n}$ is the usual line bundle over $\P^1$, we can define the following basis on $\PS$ \cite{Mason:2005zm}:
\begin{subequations}\label{eq:basis}
\begin{align}
  (0,1)\text{-vectors}\quad &:\quad  &\bar{\partial}_0&=\langle \lambda\hat{\lambda}\rangle \lambda_{\alpha}\frac{\partial}{\partial \hat{\lambda}_{\alpha}}\,,\quad  &\bar{\partial}_{\dot\alpha}&=-\lambda^{\alpha}\partial_{\alpha\dot\alpha}\,,\\
   (0,1)\text{-forms}\quad &:\quad  &\bar{e}^0&=\frac{\langle \hat{\lambda}d\hat{\lambda}\rangle}{\langle \lambda \hat{\lambda}\rangle^2}\,,\qquad \qquad  &\bar{e}^{\dot\alpha}&=-\frac{\hat{\lambda}_{\alpha}dx^{\alpha\dot\alpha}}{\langle \lambda\hat{\lambda}\rangle}\,,
\end{align}
\end{subequations}
Since $\bar{\partial}^2=0$ where $\bar{\partial}:=\bar{e}^0\bar{\partial}_0+\bar{e}^{\dot\alpha}\bar{\partial}_{\dot\alpha}$, we can take $\bar{\partial}$ to be our definition of integrable complex structure on $\PS$. Strictly speaking, $\bar{\pl}_0$ and $\bar{\pl}_{\dot\alpha}$ are $(0,1)$-vector fields of $\Gamma(T^{0,1}\PT,\cO(2))$ and $\Gamma(T^{0,1}\PT,\cO(1))$, respectively. Here, the $\Gamma$ notation is an abbreviation of the set of $C^{\infty}$ sections valued in $\cO(n)$ line bundle.

To obtain an explicit expression for $F^{\dot\alpha}(x,\lambda)$, we recall that $(A)dS_4$ with the metric \eqref{eq:metric} is described by the following system of equations \cite{Bolotin:1999fa}
\begin{subequations}
\begin{align}
    0&=de^{\alpha\dot\alpha}-\varpi^{\alpha}{}_{\gamma}\wedge e^{\gamma\dot\alpha}-\varpi^{\dot\alpha}{}_{\dot\gamma}\wedge e^{\alpha\dot\gamma}\,,\\
    0&=d\varpi^{\alpha\beta}-\varpi^{\alpha}{}_{\gamma}\wedge\varpi^{\beta\gamma}-\Lambda e^{\alpha}{}_{\dot\gamma}\wedge e^{\beta\dot\gamma}\,,\\
    0&=d\varpi^{\dot\alpha\dot\beta}-\varpi^{\dot\alpha}{}_{\dot\gamma}\wedge \varpi^{\dot\beta\dot\gamma}-\Lambda e_{\gamma}{}^{\dot\alpha}\wedge e^{\gamma\dot\beta}\,,
\end{align}
\end{subequations}
where $e^{\alpha\dot\alpha}=\Omega \sigma^{\alpha\dot\alpha}$ and $\sigma^{\alpha\dot\alpha}$ are the Pauli's matrices. The spin-connections $\varpi$ read
\begin{align}
    \varpi^{\alpha\alpha}=\Lambda \Omega \sigma^{\alpha\dot\gamma}x^{\alpha}{}_{\dot\gamma}\,,\qquad  \varpi^{\dot\alpha\dot\alpha}=\Lambda \Omega \sigma^{\gamma\dot\alpha}x_{\gamma}{}^{\dot\alpha}\,.
\end{align}
The connection $\nabla=d+\varpi$ on $\cM$ is defined as 
\begin{align}\label{spacetimecovariantder}
    \nabla A^{\alpha,\dot\alpha}&=d A^{\alpha,\dot\alpha}+\varpi^{\alpha}_{\ \beta}A^{\beta,\dot\alpha}+\varpi^{\dot\alpha}_{\ \dot\beta}A^{\alpha,\dot\beta}\,.
\end{align}
The corresponding spin-connection on $\PS$ is then a $(0,1)$-form $\bar{\varpi}=\varpi_0\bar{e}^0+\varpi_{\dot\alpha}\bar{e}^{\dot\alpha}$ that has the following property $\varpi_{0}\bar{e}^0|_{L_x}\in H^{0,1}(L_x,\cO(2))=0$ \cite{Eastwood:1981jy,Woodhouse:1985id}. This allows us to define a background connection as
\begin{align}
    \bar{\partial}_{\dot\alpha}\rightarrow \bar{\nabla}_{\dot\alpha}:=-\lambda^{\alpha}\nabla_{\alpha\dot\alpha}=\bar{\partial}_{\dot\alpha}-\lambda^{\alpha}\varpi_{\alpha\dot\alpha}\,,
\end{align}
where $\nabla_{\alpha\dot\alpha}$ is the covariant derivative defined in \eqref{spacetimecovariantder}. At the end of the day, the equations that we use to solve for the incident relations $\mu^{\dot\alpha}=F^{\dot\alpha}(x,\lambda)$ are
\begin{align}
    \lambda^{\alpha}\nabla_{\alpha\dot\alpha}\mu^{\dot\beta}=0\,.
\end{align}
We obtain
\begin{align}\label{eq:increl}
\mu^{\dot\alpha}=x^{\alpha\dot\alpha}\,\lambda_{\alpha}\quad \Leftrightarrow\quad  x^{\alpha\dot\alpha}=\frac{\lambda^{\alpha}\hat{\mu}^{\dot\alpha}-\hat{\lambda}^{\alpha}\mu^{\dot\alpha}}{\langle \lambda \hat{\lambda}\rangle}\,.
\end{align}
Thus, each point $x\in\cM$ corresponds to a holomorphic, linearly embedded Riemann sphere $L_x\cong\P^1\subset\PT$, and any point $Z\in\PT$ corresponds to a self-dual null $\alpha$-plane in $\cM$. Note that if we consider
\begin{align}\label{eq:S7}
    N=I_{AB}Z^A\hat Z^B=\langle \lambda \hat\lambda\rangle  +\Lambda [\mu\hat\mu]\,,\qquad \qquad N\in \R^+\,,
\end{align}
then a straightforward computation leads to
\begin{align}
    x^2=\frac{1}{2}x_{\alpha\dot\alpha}x^{\alpha\dot\alpha}=\frac{[\mu\hat\mu]}{\langle \lambda\hat\lambda\rangle}\,
\end{align}
in empty $(A)dS_4$. Therefore, we can identify 
\begin{align}
    \langle \lambda \hat\lambda\rangle = N\Omega\,,\qquad \qquad [\mu\hat\mu]=N\Omega\,x^2\,,
\end{align}
where $\Omega$ is the conformal factor in the metric \eqref{eq:metric}. From this point of view, the inner product $\langle \lambda\hat\lambda\rangle$ can be thought of as the conformal factor $\Omega$ as observed in \cite{Steinacker:2022jjv}. This suggests us to parametrize $\lambda,\hat\lambda$ projectively as
\begin{align}
    \lambda=\Omega^{1/2}\binom{ze^{i\theta}}{-1}\,,\qquad \qquad \hat\lambda=\Omega^{1/2}\binom{1}{ze^{-i\theta}}\,,
\end{align}
where $|z|^2+1=N$ and $z\in \R^*$, $\theta\in [0,2\pi]$. Next, plugging \eqref{eq:increl} to the basis \eqref{eq:basis}, we recover the usual definition of the Dolbeault operator on $\PT$, i.e.
\begin{align}
    \bar{\partial}=d\lambda^{\alpha}\frac{\partial}{\partial \lambda^{\alpha}}+d\mu^{\dot\alpha}\frac{\partial}{\partial \mu^{\dot\alpha}}=d\hat Z^A\frac{\partial}{\partial \hat Z^A}\,.
\end{align}
Hence, $\bar{\pl}$ can play the role of the background on the twistor space $\PT$ associated to complex conformally flat spacetime $\cM$. For this reason, we will abusively denote $\bar{\nabla}$ also as $\bar{\pl}$, where $\bar{\nabla}$ is the corresponding connection of $\nabla$ on $\PS$.

\paragraph{Deformation of twistor geometry.} In the study of deformation of complex structures on twistor space \cite{Penrose:1972ia,Ward:1977ta}, one often deforms the ``background'' $\bar{\pl}$ by some connection $(0,1)$-form $\sa$ with homogeneity zero on $\PTc$, i.e.
\begin{align}\label{eq:deformationofcomplexstructures}
    \bar{\pl}\mapsto \bar{\Dc}=\bar{\pl}+\sa\,,\qquad \qquad \sa\in \Omega^{0,1}(\PTc,\cO)\,.
\end{align}
The integrability condition for the deformed complex structure $\bar{\Dc}$ is the Kodaira-Spencer equation:
\begin{align}\label{eq:Kodaira-Spencer}
    \sF:=\bar{\Dc}\wedge \bar{\Dc}=\bar\pl \sa +\sa\wedge \sa=0\,,\qquad  \qquad \sF\in \Omega^{0,2}(\PTc)\,,
\end{align}
In this situation, the deformed twistor space $\PTc$ will corresponds to a self-dual background associated with the deformation $\sa$, and is defined as
\begin{align}\label{eq:deformedPT}
    \PTc=\big\{\cZ^A=(\lambda^{\alpha},\mu^{\dot\alpha}=F^{\dot\alpha}(x,\lambda))\in \P^3\,\big|\,\bar{\pl}F^{\dot\alpha}|_{L_x}=\sa^{\dot\alpha}|_{L_x}\big\}\,.
\end{align}
While the twistor lines $L_x\subset \PTc$ are sections of $\pi:\PTc\rightarrow \P^1$ with the same normal bundle $\cO(1)\oplus \cO(1)$ as before, deforming $\bar{\pl}$ leads to a distortion of $L_x$ away from the original twistor line $x^{\alpha\dot\alpha}\lambda_{\alpha}$ by a displacement $(F^{\dot\alpha}-x^{\alpha\dot\alpha}\lambda_{\alpha})\frac{\pl}{\pl \mu^{\dot\alpha}}$. This deformation can be depicted by the following cartoon:
\begin{figure}[h!]
    \centering
    \includegraphics[scale=0.53]{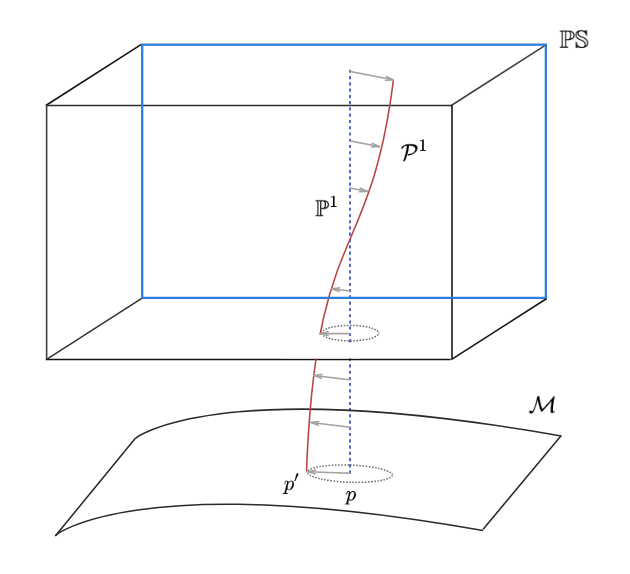}
\end{figure}

As shown in \cite{kodaira1962theorem}, the moduli space of solutions to the PDE \eqref{eq:deformedPT} has complex dimensions 4, and is identified with a self-dual background $\Mcurl=\cM\oplus (\text{deformations})$. Furthermore, the incident relations are no longer of the form \eqref{eq:increl} but rather a solution of
\begin{align}
    \lambda^{\alpha}\tilde\nabla_{\alpha\dot\alpha}F^{\dot\beta}(x,\lambda)=0\,.
\end{align}
where $\tilde \nabla$ is a self-dual connection on $\cM$. The operator $\ellbold_{\dot\alpha}:=\lambda^{\alpha}\tilde\nabla_{\alpha\dot\alpha}$ is known as Lax pair whose integrability, i.e. $[\ellbold_{\dot\alpha},\ellbold_{\dot{\beta}}]=0$, is equivalent to the self-dual vacuum equations on $\Mcurl$. Note that depends on the nature of $\sa$, we have either: (i) a radiative self-dual background \cite{Ward:1977ta}, (ii) gravitational self-dual background \cite{Penrose:1972ia}, (iii) higher-spin extension of self-dual radiative background \cite{Tran:2021ukl}, etc. Furthermore, in the radiative cases, the incident relations reduce to the usual ones in \eqref{eq:increl}, while it is more complicated for higher-derivative deformations, see e.g. \cite{Penrose:1972ia,Adamo:2021bej}.

To this end, let us introduce a Poisson structure induced by the infinity twistor $I^{AB}$ on $\PTc$:
\begin{align}
    \widetilde{\Pi}=I^{AB}\frac{\overleftarrow{\partial}}{\pl \cZ^A}\wedge \frac{\overrightarrow{\partial}}{\pl \cZ^B}=\Lambda \frac{\overleftarrow{\partial}}{\partial \lambda^{\alpha}}\wedge \frac{\overrightarrow{\partial}}{\partial \lambda_{\alpha}}+\frac{\overleftarrow{\partial}}{\partial \mu^{\dot\alpha}}\wedge \frac{\overrightarrow{\partial}}{\partial \mu_{\dot\alpha}}\,.
\end{align}
A nice feature of the above Poisson structure is that it can act naturally on holomorphic objects on curved twistor space. Furthermore, it also induces the following star-product on $\PTc$ \cite{Haehnel:2016mlb,Adamo:2016ple,Tran:2022tft}
\begin{align}
    f\star g&:=fe^{\ell_p \widetilde\Pi}\wedge g=\sum_{k=0}^{\infty}\frac{\ell_p^k}{k!}f\,\widetilde\Pi^k\,g\,,\label{eq:starproduct}
\end{align}
where $\ell_p$ is some natural length scale that plays the role of a deformation parameter. Note that the expansion of the $\star$-product at order $k$ has weight $-2k$ for $k \in \N$. This allows us to obtain the $(+++)$ cubic vertices in chiral HSGRA on twistor space since all twistor fields at the cubic vertex can have positive weights. 
We also note that $(A)dS_4$ can be realized in terms of the following generators
\begin{align}
    L^{\alpha\beta}=\lambda^{\alpha}\lambda^{\beta}\,,\qquad \qquad P^{\alpha\dot\alpha}=\lambda^{\alpha}\mu^{\dot\alpha}\,,\qquad \qquad L^{\dot\alpha\dot\beta}=\mu^{\dot\alpha}\mu^{\dot\beta}\,,
\end{align}
where due to the definition of the $\star$-product, we have 
\begin{subequations}
\begin{align}
    [L^{\alpha\alpha},L^{\beta\beta}]_{\star}&=\Lambda\eps^{\alpha\beta}L^{\alpha\beta}\,, && &[L^{\dot\alpha\dot\alpha},L^{\dot\beta\dot\beta}]_{\star}&=\eps^{\dot\alpha\dot\beta}L^{\dot\alpha\dot\beta}\,,\\
    [L^{\alpha\alpha},P^{\beta\dot\beta}]_{\star}&=\Lambda \eps^{\alpha\beta}P^{\alpha\dot\beta}\,, && &[L^{\dot\alpha\dot\alpha},P^{\beta\dot\beta}]_{\star}&=\eps^{\dot\alpha\dot\beta}P^{\beta\dot\alpha}\,,\\
    [P^{\alpha\dot\alpha},P^{\beta\dot\beta}]_{\star}&=\Lambda\eps^{\alpha\beta}L^{\dot\alpha\dot\beta}+\eps^{\dot\alpha\dot\beta}L^{\alpha\beta}\,.
\end{align}
\end{subequations}
Intriguingly, in the flat limit, only `half' of the above relations remain. In addition, the above realization of $(A)dS_4$ algebra does not reduce to the usual Poincare algebra in flat space but rather a Maxwell algebra \cite{Bacry:1970ye,Schrader:1972zd,Bonanos:2008ez,Ponomarev:2022ryp} that describes self-dual spacetime.

\paragraph{The twistor action for chiral HSGRA.} Since we are working with $\PTc$, everything must have appropriate projective scalings. Therefore, it is useful to define the following Euler operator
\begin{align}\label{Euler1}
    \cE=\cZ^A\frac{\partial}{\partial \cZ^A}\,,
\end{align}
to keep track of the homogeneity in $\cZ$ for any twistor expression. 

It is well-known that the canonical bundle of $\P^3$ is $\cO(-4)$. This fact motivates us to choose the following $SU(4)$-invariant measure 
\begin{align}\label{eq:measurePTc}
    D^3\cZ=\epsilon_{ABCD}\cZ^Ad\cZ^B\wedge d\cZ^C\wedge d\cZ^D=\langle \lambda d\lambda\rangle \wedge [d\mu\wedge d\mu]\,
\end{align}
as the canonical measure on $\PTc$ where $\cE (D^3\cZ)=4$. The twistor action for chiral HSGRA in (anti)de-Sitter space which admits a smooth flat limit is \cite{Tran:2022tft}
\begin{align}\label{eq:Chern-Simonsaction}
\begin{split}
    S[\A]&=S_{hCS}+S_c=\int D^3\cZ \,\text{Tr}\big[\sum_{h\in \Z}\A_{-h} \star \bar{\partial}\A_h + \frac{2}{3}\sum_{h_i\in \Z}\A_{h_1} \star \A_{h_2} \star \A_{h_3}\big]+S_c\,,
    \end{split}
\end{align}
where the twistor field $\A_h\in\Omega^{0,1}(\PTc,\text{End}(E)\otimes\cO(2h-2))$ corresponds to a matrix-valued higher-spin fields of helicity $h$ in spacetime. Furthermore, we must have at least one positive-helicity field in \eqref{eq:Chern-Simonsaction} so that the constraint $\cE \,(L[\A])=-4$ can be fulfilled. Here, $L[\A]$ is the Lagrangian associated with the action \eqref{eq:Chern-Simonsaction}. Note that the measure \eqref{eq:measurePTc} is not gauge-invariant under the higher-spin diffeomorphism:
\begin{align}\label{diffeo}
    \delta \cZ^A=\sum_{h\in \Z}\{\cZ^A,\xi_h\}=\sum_{h\in \Z}(\Lambda \pl^{\alpha}\xi_h+\pl^{\dot\alpha}\xi_h)\,,\qquad \xi_h\in \Gamma(\PTc,\cO(2h-2))\,.
\end{align}
For this reason, a correction denoted as $S_c$ has to be added to the above action (c.f. \eqref{eq:Chern-Simonsaction}).
\paragraph{Scattering amplitudes.} Remarkably, by doing integration by parts, the number of the star-products in each term of the action $S_{hCS}$ can be reduced by one \cite{Tran:2022tft}. Accordingly, in finding amplitudes from twistor space, it is convenient to choose the following twistor representative for momentum eigenstates \cite{Haehnel:2016mlb,Adamo:2016ple}
\begin{align}
\begin{split}
    \A_{h_i}&=\int \frac{dt_i}{t_i^{2h_i-1}}\bar{\delta}^2(t_i\lambda-\lambda_{i})e^{t_i[\mu\tilde{\lambda}_i]}\,,\qquad h_i\in \Z\,,
\end{split}
\end{align}
in terms of the on-shell four-momentum $k_i^{\alpha\dot\alpha}=\lambda_i^{\alpha}\tilde\lambda_i^{\dot\alpha}$, which is a null vector on the tangent space of $(A)dS_4$. Here,
\begin{align}
    \bar{\delta}(az-b)=\frac{1}{2\pi i}d\bar{z}\frac{\partial}{\partial \bar{z}}\Big(\frac{1}{az-b}\Big)\,,
\end{align}
is a $(0,1)$-form holomorphic delta function \cite{Witten:2004cp}. If we consider $\lambda_{\alpha}=(1,z)$ and $\lambda_{\alpha}'=(b,a)$, the above can be recast into:
\begin{align}
    \bar{\delta}(\langle\lambda\lambda'\rangle)=\frac{1}{2\pi i}d\overline{\lambda^{\dot\alpha}}\frac{\partial}{\partial \overline{\lambda^{\dot\alpha}}}\frac{1}{\langle \lambda\lambda'\rangle}
\end{align} 
In addition, we can define a projective version of the holomorphic delta function:
\begin{align}
    \bar{\delta}_m(\lambda,\lambda')=\Big[\frac{\langle \xi \lambda\rangle}{\langle \xi\lambda'\rangle}\Big]^m\bar{\delta}(\langle\lambda\lambda'\rangle)=\int_{\mathbb{C}}\frac{dt}{t^m}\bar{\delta}^2(t\lambda-\lambda')
\end{align}
for $(0,1)$-form `currents' that can have different scaling. As shown in \cite{Nagaraj:2018nxq,Nagaraj:2019zmk}, the plane wave solutions in the affine patch of $(A)dS_4$ have the same structures with the ones in flat space. This explains why the above above twistor representative can be chosen. We arrive at (up to some overall factor)
\begin{equation}
    \begin{split}
    \cM_3^{\Lambda}(h_1,h_2,h_3)=&\frac{\ell_p^k}{k!}\int d^2\lambda \,dt_1dt_2dt_3\,t_1^{1-2h_1}t_2^{k+1-2h_2}t_3^{k+1-2h_3}\left([23]+\Lambda\Big\langle\frac{\partial}{\partial \lambda_2}\frac{\partial}{\partial \lambda_3}\Big\rangle\right)^k \\
    &\times \bar{\delta}^2(t_1\tilde\lambda_1+t_2\tilde\lambda_2+t_3\tilde\lambda_3)\bar{\delta}^2(t_1\lambda-\lambda_1)\bar{\delta}^2(t_2\lambda-\lambda_2)\bar{\delta}^2(t_3\lambda-\lambda_3)\,.
    \end{split}
\end{equation}
The three-point amplitudes read
\begin{align}\label{3ptAdS}
    \begin{split}
    \cM_3^{\Lambda}(h_1,h_2,h_3)
    =[12]^{h_1+h_2-h_3}[23]^{h_2+h_3-h_1}[31]^{h_3+h_1-h_2}\frac{\big[\ell_p(1-\Lambda \Box_P)\big]^{h_1+h_2+h_3-1}}{\Gamma[h_1+h_2+h_3]}\delta^4(P)\,.
    \end{split}
\end{align}
From this, we can extract the celebrated coupling constants of chiral HSGRA 
\begin{align}\label{eq:couplingconstants}
    \bar{\cC}_{h_1,h_2,h_3}=\frac{\ell_p^{h_1+h_2+h_3-1}}{\Gamma[h_1+h_2+h_3]}\,,\qquad \quad  h_1+h_2+h_3>0\,.
\end{align}
Note that in our construction, the coupling constant $\bar{\cC}_{h_1,h_2,h_3}$ naturally comes from the definition of the $\star$-product.

\paragraph{Reduction to spacetime.} While obtaining the scattering amplitudes from twistor space is a reasonably simple task, integrating out fibre coordinates is quite complicated. The reason is that we have to take into account what is so-called ``frame-dragging'' effects on $\PS$ since the pullback $\pi_1^*:\PTc\rightarrow\PS$ is not holomorphic. To illustrate how this subtle effect affects the process of integrating out fibre coordinates, it is enough for us to consider a $\PTc$ associated with a flat target space. To begin, let us define the following $(1,0)$-vector fields and their dual $(1,0)$-forms:
\begin{subequations}\label{eq:nicebasisfordeformation}
\begin{align}
 (1,0)\text{-vectors} &:  &\partial_0&:=\frac{\hat{\lambda}_{\alpha}}{\langle \lambda \hat \lambda \rangle}\frac{\partial}{\partial \lambda_{\alpha}}\,, && &\partial_{\dot\alpha}&:=-\frac{\hat{\lambda}^{\alpha}}{\langle \lambda\hat\lambda\rangle}\partial_{\alpha\dot\alpha}\label{eq:10basis}\,,\\
  (1,0)\text{-forms} &:  &e^0&:=\langle \lambda d\lambda\rangle\,, && &e^{\dot\alpha}&:=\lambda_{\alpha}dx^{\alpha\dot\alpha}\,.
\end{align}
\end{subequations}
where $\pl_0$ and $\pl_{\dot\alpha}$ are elements of $\Gamma(T^{1,0}\PT,\cO(-2))$ and $\Gamma(T^{1,0}\PT,\cO(-1))$, respectively. Using \eqref{eq:basis} and \eqref{eq:nicebasisfordeformation}, we can check that
\begin{align}
    [\bar{\pl}_0,\pl_{\dot\alpha}]=\bar{\pl}_{\dot\alpha}\,,\qquad \qquad [\bar{\pl}_{\dot\alpha},\pl_0]=\pl_{\dot\alpha}\,.
\end{align}

Note that when $\pl_{\dot\alpha}=\frac{\pl}{\pl \mu^{\dot\alpha}}$ acts on functions, it is sufficient to use the definition of $\pl_{\dot\alpha}$ in \eqref{eq:10basis} when we pull it back from $\PTc$ to $\PS$. However, when the $\pl_{\dot\alpha}$-vector field acts on differential forms, we must think of this operation as a Lie derivative generated by a flow along $\pl_{\dot\alpha}$. This fact motivates us to introduce the following Poisson structure \cite{Aschieri:2009qh}:
\begin{align}\label{eq:holomorphicPoisson}
    \omega\,\Pi\,\eta:=\{\omega,\eta\}_h=\eps^{\dot\alpha\dot\beta}\cL_{\pl_{\dot\alpha}} \omega \wedge \cL_{\pl_{\dot\beta}}\eta\,,
\end{align}
that acts any $(p,q)$-forms $\omega,\eta$ on $\PS$ to work with higher-spin extension of Penrose transform. The holomorphic Poisson structure \eqref{eq:holomorphicPoisson} naturally induces the following star-product 
\begin{align}\label{eq:astproduct}
    \omega \ast \eta :=\omega\, e^{\ell_p \Pi}\wedge \eta =\sum_{k=0}^{\infty}\frac{\ell_p^k}{k!}\omega \,\Pi^k\,\eta\,,
\end{align}
where $\omega$ and $\eta$ are differential forms. 

Instead of working directly with $\PS$, it will be more convenient to work with the non-projective undotted spinor bundle $\mathbb S$ since we can skip the step of trivializing line bundles in different patches of $\PS$, i.e. we do not need to find a holomorphic frame $\e$ where 
\begin{align}
    \bar{\Dc}\big|_{L_x}\e=0\,.
\end{align}
The tradeoff is that we need to project all functions/forms from $\mathbb S$ to $\PS$ appropriately with the help of the following Euler operator 
\begin{align}
    \hatSigma=\lambda^{\alpha}\frac{\pl}{\pl \lambda^{\alpha}}\,
\end{align}
on $\C^2\backslash\{0\}$ to pick out the correct projective scaling. Any $\omega\in \Omega^{p,q}(\PTc,\cO(n))$ is demanded to satisfy
\begin{align}
    \hatSigma\intprod \omega=0\,,\qquad \qquad \cL_{\hatSigma}\omega=n\,\omega\,.
\end{align}
Here, the notation $\hatSigma\intprod \equiv \iota_{\hatSigma}$ is the interior product wrt. to the $\hatSigma$ vector field. Furthermore, $L_x\simeq \P^1\sim S^2$ is endowed with a hermitian Fubini-Study metric of the form
\begin{align}
   \tK:= e^0\wedge \bar{e}^0=\frac{\langle \lambda d\lambda\rangle \wedge\langle \hat \lambda d\hat\lambda\rangle }{\langle \lambda\hat\lambda\rangle^2}\,.
\end{align}
The usual exterior derivative on $\PS$, denoted as $d_{\PS}\equiv \eth$, gets lifted to a unique Chern connection on the bundle $\cO(n)\rightarrow\P^1$ where \cite{Herfray:2016qvg,Sharma:2021pkl}
\begin{align}
    d_{\PS}\equiv \eth:=d_{\mathbb S}+n\frac{\langle \hat \lambda d\lambda\rangle}{\langle \lambda \hat\lambda\rangle}\wedge\,.
\end{align}
Here,
\begin{align}
    d_{\mathbb S}:=e^0\pl_{0}+\bar e^0\bar{\pl}_0+dx^{\alpha\dot\alpha}\frac{\pl}{\pl x^{\alpha\dot\alpha}}=d\lambda^{\alpha}\frac{\pl}{\pl \lambda^{\alpha}}+d\hat\lambda^{\alpha}\frac{\pl}{\pl 
    \hat\lambda^{\alpha}}+dx^{\alpha\dot\alpha}\frac{\pl}{\pl x^{\alpha\dot\alpha}}
\end{align}
is the exterior derivative on the unprojective spinor bundle $\mathbb S$. The Lie derivative $\cL_{\pl_{\dot\alpha}}$ acting on $\cO(n)$-valued $(p,q)$-forms can be defined via the Cartan's magic formula by
\begin{align}
    \cL_{\pl_{\dot\alpha}}\omega=\pl_{\dot\alpha}\intprod \eth \omega+\eth (\pl_{\dot\alpha}\intprod \omega)\,.
\end{align}
It is easy to check that 
\begin{align}\label{eq:dragging}
    \eth e^0=\eth \bar{e}^0=0\,,\qquad \eth e^{\dot\alpha}=e^0\wedge \bar{e}^{\dot\alpha}\,,\qquad \eth \bar{e}^{\dot\alpha}=e^{\dot\alpha}\wedge \bar{e}^0\,.
\end{align}
Furthermore, since $\eth:= \pl+\bar{\pl}$, we find
\begin{align}
    \pl \bar{e}^0=0\,,\qquad \pl \bar{e}^{\dot\alpha}=e^{\dot\alpha}\wedge \bar{e}^0\,.
\end{align}
This is what is so-called the ``frame-dragging'' effect that occurs on twistor space whenever we deform twistor space with a potential that contains derivatives along the fibres $\cO(1)\oplus \cO(1)$ of the fibration $\pi:\PTc\rightarrow L_x$. Let us now focus on holomorphic form on $\PS$ where we can decompose $\A=\A_0\bar{e}^0+\A_{\dot\alpha}\bar{e}^{\dot\alpha}$. A simple computation provides
\begin{subequations}
\begin{align}
    \bar{\pl}\A&=(\bar{\pl}_0\A_{\dot\alpha}-\bar{\pl}_{\dot\alpha}\A_0)\bar e^0\wedge \bar e^{\dot\alpha}+\bar{\pl}_{\dot\alpha}\A_{\dot\beta}\bar{e}^{\dot\alpha}\wedge \bar{e}^{\dot\beta}\,,\\
    \pl\A&=\pl_0\A_0e^0\wedge \bar{e}^{0}+\pl_0\A_{\dot\alpha}e^0\wedge \bar{e}^{\dot\alpha}-(\pl_{\dot\alpha}\A_0+\A_{\dot\alpha})\bar{e}^0\wedge e^{\dot\alpha}+\pl_{\dot\alpha}\A_{\dot\beta}e^{\dot\alpha}\wedge \bar{e}^{\dot\beta}\,,
\end{align}
\end{subequations}
which can be used to obtain
\begin{align}
    \cL_{\pl_{\dot\alpha}}\A=(\pl_{\dot\alpha}\A_0+\A_{\dot\alpha})\bar{e}^0+\pl_{\dot\alpha}\A_{\dot\beta}\bar{e}^{\dot\beta}\,,
\end{align}
where $\pl_{\dot\alpha}\intprod \A=0$; and vector fields `eat' differential forms according to the following menu:
\begin{align}
    \pl_0\intprod e^0=1\,,\quad \bar{\pl}_0\intprod\bar{e}^0=1\,,\qquad \pl_{\dot\alpha}\intprod e^{\dot\beta}=\delta_{\dot\alpha}{}^{\dot\beta}\,,\quad \bar{\pl}_{\dot\alpha}\intprod\bar{e}^{\dot\beta}=\delta_{\dot\alpha}{}^{\dot\beta}\,.
\end{align}
Due to the holomorphicity of $\A$, we can show that
\begin{align}
    \cL_{\pl_{\dot\alpha_1}}...\cL_{\pl_{\dot\alpha_k}}\A=(\pl_{\dot\alpha_1}...\pl_{\dot\alpha_k}\A_0+k\pl_{(\dot\alpha_1}...\pl_{\dot\alpha_{k-1}}\A_{\dot\alpha_k)})\bar{e}^0+\pl_{\dot\alpha_1}...\pl_{\dot\alpha_k}\A_{\dot\beta}\bar{e}^{\dot\beta}\,.
\end{align}
We will choose the following gauge condition \cite{Sharma:2021pkl} 
\begin{align}\label{eq:gaugechoice}
    \cL_{\partial_{\dot\alpha}}\A^{\dot\alpha}=\partial_{\dot\alpha}\A^{\dot\alpha}=0\,,
\end{align}
to reduce the number of $\ast$-product by one. Indeed, one can show that $\cL_{\pl_{\dot\alpha}}\cL_{\pl^{\dot\alpha}}\A=0$ if \eqref{eq:gaugechoice} holds. Once again, we have only one derivative, i.e. $\bar{\partial}$, in the kinetic term. Pulling \eqref{eq:Chern-Simonsaction} back to $\PS$, we get
\begin{align}\label{eq:actionCS}
    S[\A]=\int_{\PS} \Tr\Big[\sum_{h\in \mathbb{Z}}\A_{-h} \bar{\partial}\A_h + \frac{2}{3}\sum_{h_i\in \mathbb{Z}}\A_{h_1} \A_{h_2} \ast \A_{h_3}\Big]\,.
\end{align}
With a little more effort, we end up at
\begin{align}
    \begin{split}
    &\cL_{\pl_{\dot\alpha_1}}...\cL_{\pl_{\dot\alpha_k}}\A \wedge \cL_{\pl^{\dot\alpha_1}}...\cL_{\pl^{\dot\alpha_k}}\A\\
    &=(\pl_{\dot\alpha_1}...\pl_{\dot\alpha_k}\A_0+k\pl_{(\dot\alpha_1}...\pl_{\dot\alpha_{k-1}}\A_{\dot\alpha_k)})\pl^{\dot\alpha_1}...\pl^{\dot\alpha_k}\A_{\dot\beta}\bar{e}^0 \bar{e}^{\dot\beta}+\pl_{\dot\alpha_1}...\pl_{\dot\alpha_k}\A_{\dot\beta}\pl^{\dot\alpha_1}...\pl^{\dot\alpha_k}\A_{\dot\gamma}\bar{e}^{\dot\beta} \bar e^{\dot\gamma}\,.
    \end{split}
\end{align}
Observe that Woodhouse gauge (fixing $\A_0\in \Omega^{0,1}(\P^1,\cO(n)) = 0$ for $n\geq -1$ \cite{Woodhouse:1985id}) is no longer a suitable gauge choice since there is an in-homogeneous contribution to the $\bar{e}^0\wedge \bar{e}^{\dot\alpha}$ component of the equation of motion
\begin{align}
    \bar\pl \A+\A\ast \A=0\,.
\end{align}
In terms of components, the above decomposes into two sub-equations:
\begin{subequations}
\begin{align}
    0&=\bar\pl_0\A_{\dot\alpha}+(\pl_{\dot\beta_1}...\pl_{\dot\beta_k}\A_0+k\pl_{(\dot\beta_1}...\pl_{\dot\beta_{k-1}}\A_{\dot\beta_k)})\pl^{\dot\beta_1}...\pl^{\dot\beta_k}\A_{\dot\alpha}\,,\label{eq:bare0eda}\\
    0&=\bar\pl_{\dot\alpha}\A_{\dot\beta}+\pl_{\dot\gamma_1}...\pl_{\dot\gamma_k}\A_{\dot\alpha}\pl^{\dot\gamma_1}...\pl^{\dot\gamma_k}\A_{\dot\beta}\,.
\end{align}
\end{subequations}
It can be shown that if we assume $\A_{\dot\alpha}$ to have the usual Woodhouse representative for positive-helicity higher-spin fields, then the solution of \eqref{eq:bare0eda} is 
\begin{align}\label{eq:A0step1}
    \partial_{\dot\alpha}\A_0=-\A_{\dot\alpha}\,,
\end{align}
This implies that $\A_{\dot\alpha},\A_0$ are holomorphic in $\lambda$ when they have positive weight since $\bar{\pl}_0\pl_{\dot\alpha}=-\bar{\pl}_{\dot\alpha}$. As a result, we have the following equations
\begin{align}
    \bar{\pl}_0\A_{\dot\alpha}=0\,,\qquad \qquad \bar{\pl}_0\A_0=0\,
\end{align}
that address the holomorphicity of $\A_0,\A_{\dot\alpha}\in \Omega^{0,1}(\PTc,\cO(n))$ when $n\geq 0$. If we consider the dotted component of $\A$ with the usual Woodhouse representative \cite{Woodhouse:1985id}
\begin{align}\label{eq:generatingfunction}
    \begin{split}
    \A_{h,\dot\alpha}\bar{e}^{\dot\alpha}=\lambda^{\alpha(2h-1)}A_{\alpha(2h-1),\dot\alpha}\,\bar{e}^{\dot\alpha}\,,
    \end{split}
\end{align}
then the zero component of $\A$ takes the following form \cite{Tran:2022tft}
\begin{subequations}
\begin{align}
    \A_{h,0}^+\bar{e}^0&= -\frac{\bar{\partial}^{\dot\alpha}}{\Box} \A_{\dot\alpha}\bar e^0\,,\qquad\quad  h>0\,,\label{eq:dangerousBox}\\
    \A_{h,0}^-\bar{e}^0&=\frac{\hat{\lambda}_{\alpha(2|h|)}}{\langle \lambda \hat\lambda\rangle^{2|h|}}B^{\alpha(2|h|)}\bar{e}^0\,,\quad \quad   h\leq 0\,,
\end{align}
\end{subequations}
where we used the convention $\lambda^{\alpha(s)}=\lambda^{(\alpha_1}...\lambda^{\alpha_s)}$ etc. to shorten our expressions. As always, the scalar field is the most problematic piece to be inserted into the spectrum of HSGRAs, given that it has to respect higher-spin symmetry. Let us consider the following twistor field
\begin{align}
   \A_{h=0^+}:= \vartheta=\frac{\hat{\lambda}_{\alpha}}{\langle \lambda\hat\lambda\rangle}\vartheta^{\alpha}{}_{\dot\alpha}\bar{e}^{\dot\alpha}\,.
\end{align}
Here, $\vartheta^{\alpha}{}_{\dot\alpha}$ is the auxiliary field associated to the scalar field, which can be integrated out by its own equation of motion as observed in \cite{Boels:2006ir}. As a straightfoward exercise, we can show that
\begin{align}\label{eq:varthetaexplicit}
    \vartheta^{\alpha}{}_{\dot\alpha}=\Big(\partial^{\alpha}{}_{\dot\alpha}+\A^{\alpha}{}_{\dot\alpha}\ast\Big)\A_0+\frac{1}{k!}\partial_{\dot\gamma_1}...\partial_{\dot\gamma_{k+1}}\A_{\dot\alpha}\pl^{(\dot\gamma_1}...\pl^{\dot\gamma_k}\A^{\dot\gamma_{k+1})}\,,
\end{align}
where $\A_{\alpha\dot\alpha}\in\{ \oplus_s\Gamma(\PT,\text{End}(E)\otimes\cO(2s-2))\,|\,s\geq 1\}\,$. With this information, we can write \eqref{eq:actionCS} as
\small
\begin{align}
    \begin{split}
    S=\int_{\PS} \mho\,\Tr\Big[\A_0(\bar{\partial}_{\dot\alpha}+\A_{\dot\alpha}\ast )\A^{\dot\alpha}+\frac{\lambda^{\gamma}\hat\lambda^{\beta}}{\langle \lambda\hat\lambda\rangle}\vartheta_{\alpha\dot\alpha}\,\vartheta_{\beta}{}^{\dot\alpha}\Big]+S_c\,,
    \end{split}
\end{align}
\normalsize
where the measure $\mho$ is \cite{Mason:2005zm}
\begin{align}
    \mho=D^3Z\,\bar{e}^0\wedge  [\bar{e}^{\dot\alpha}\wedge  \bar{e}_{\dot\alpha}]=d^4x\frac{\langle \lambda d\lambda\rangle\wedge \langle \hat{\lambda}d\hat{\lambda}\rangle  }{\langle \lambda \hat{\lambda}\rangle^2}=d^4x\,\tK\,,
\end{align}
and $\tK$ is the top form on $\P^1$. To obtain the spacetime action for chiral HSGRA, we can use the following integral over $\P^1$ \cite{Woodhouse:1985id,Boels:2006ir,Jiang:2008xw}:
    \begin{align}\label{eq:bridge}
    \int_{\P^1}\tK\, \frac{\hat{\lambda}_{\alpha(m)}\,\lambda^{\beta(m)}}{\langle \lambda \hat{\lambda}\rangle^{m}}=-\frac{2\pi i}{(m+1)}\epsilon^{\ \beta}_{ \alpha}...\epsilon^{\ \beta}_{ \alpha}\,.
\end{align}
The final details of this step can be found in \cite{Tran:2022tft}.

The upshot of the presented approach is that we can work with holomorphic forms and Poisson structure. As a consequence, it somewhat simplifies the structures of higher-spin interactions after integrating out fibre coordinates using \eqref{eq:bridge} as compared to the fuzzy twistor construction approach (see the next section). However, as it stands, this approach is designed mainly for self-dual theories written in terms of $BF$ or holomorphic Chern-Simons forms, see e.g. \cite{Mason:2005zm,Bittleston:2020hfv,Bittleston:2022nfr}. To go beyond self-dual sectors with this approach is rather involved and requires further technical details on Wilson loops, see e.g. \cite{Adamo:2020yzi,Adamo:2022mev}. This situation can be circumvented by using a more practical approach known as fuzzy twistor construction \cite{Steinacker:2022jjv}.
\section{HS-IKKT from fuzzy twistor space}

There is a fundamental difference between the fuzzy twistor construction, and the usual twistor construction. Namely, the primary objects of the former are functions, while they are holomorphic forms in the latter. By working directly with functions, we do not need to deal with the frame-dragging effect as before (cf. \eqref{eq:dragging}). Furthermore, everything is naturally higher-spin extensible. The only downside of this approach is that we need to have a correct action to start with. Here, we discuss how to find the higher-spin extension of the IKKT-matrix model (HS-IKKT).

\paragraph{The action.} The $SO(10)$-invariant Euclidean IKKT model has the following action functional:
\begin{align}\label{eq:IKKTaction}
    S=\Tr\Big([Y^{\boldsymbol{I}},Y^{\boldsymbol{J}}][Y_{\boldsymbol{I}},Y_{\boldsymbol{J}}]+\bar{\Psi}^{\cA}\gamma^{\boldsymbol{I}}_{\cA\cB}[Y_{\boldsymbol{I}},\Psi^{\cB}]\Big)\,,\qquad {\boldsymbol{I}}=1,...,10\,,
\end{align}
where $Y^{\boldsymbol{I}}$ are $N\times N$ hermitian matrices, and
$\Psi^{\cB}$ are matrix-valued spinors. A coordinate $y^{\boldsymbol I}$ in the target space $\R^{10}$ may be defined via localized quasi-coherent states $|y\rangle \in \cH$ as \cite{Steinacker:2020nva,Ishiki:2015saa,Schneiderbauer:2016wub,Berenstein:2012ts}:
\begin{align}
 y^{\boldsymbol{I}} = \langle y|Y^{\boldsymbol{I}}|y\rangle \ \in  \R^{10} \ ,
 \label{coherent-expect}
\end{align}
where $\cH$ is some Hilbert space. In this sense, $y^{\boldsymbol I}$ can be used to define some fuzzy manifold $\cM \hookrightarrow \R^{10}$. Thus, it is possible to associate classical functions to matrices through the map
\begin{align}
\begin{split}
{\rm Mat}(\cH) &\sim \cC(\cM) \\
\Phi &\sim \langle y|\Phi|y\rangle = \phi(y) \ .
\end{split}
\end{align}
The matrix algebra 
${\rm Mat}(\cH)$ generated by $Y^{\boldsymbol{I}}$ is interpreted as quantized algebra of functions on $\cM$.
Since $Y^{\boldsymbol I}$ is not commutative on $\cM$, it breaks Lorentz invariance. This is mitigated on covariant quantum spaces such as a fuzzy 4-sphere \cite{Sperling:2017dts} (denoted as $S^4_N$), where the non-commutativity of $Y^a\in \R^5$ is measured by a quantized Poisson structure
\begin{align}\label{eq:noncom}
    [Y^{a},Y^{b}] =: i\theta^{ab}\,,\qquad a,b=1,2,3,4,5\,.
\end{align}
Decomposing $Y^a=\bar Y^a + A^a$ where $\bar{Y}^a$ is the background $S^4_N$, the action defines a non-commutative Yang-Mills-type gauge theory on $\cM$ \cite{Aoki:1999vr}, with the gauge transformations $U^{-1}(\bar Y^a + A^a)U$.

\paragraph{Fuzzy 4-sphere in the semi-classical limit.} Consider $\R^5$ with the metric $\eta^{ab}=\diag(+,+,+,+,+)$. By requiring $Y^a$ to transform as vectors under $SO(5)$ equipped with the generators $M_{ab}$, we have the following $\mso(6)$ algebra
\begin{subequations}\label{eq:so(5)}
\begin{align}
    [M_{ab},M_{cd}]&=i(M_{ad}\delta_{bc}-M_{ac}\delta_{bd}-M_{bd}\delta_{ac}+M_{bc}\delta_{ad})\,,\\
    [M_{ab},Y_c]&=i(Y_a\delta_{bc}-Y_b\delta_{ac})\,,\\
    [Y_a,Y_b]&=i\ell_p^2 M_{ab}\,,\\
    Y_{a}Y^{a}&=R^2\,,
\end{align}
\end{subequations}
which defines a fuzzy 4-sphere with a radius $R$.

In practice, we will work mostly with almost-commutative/semi-classical limit where the natural length scale $\ell_p\simeq \frac{2R}{N}\sim 0$. In this limit, we replace \cite{Steinacker:2019fcb}
\begin{subequations}
\begin{align}
    Y^a&\rightarrow y^a\ \text{(these are commutative coordinates)}\,,\\ [\,,]&\rightarrow i\{\,,\}\,.
\end{align}
\end{subequations}
Note that the Poisson bracket $\{\,,\}$ is not the same with the holomorphic holomorphic Poisson structure defined in \eqref{eq:holomorphicPoisson}; and we shall define $\{\,,\}$ the moment we discuss about almost-non-commutative twistor space. 

\paragraph{Affine patch.} What we gain from the semi-classical limit is a proper notion of geometry. Consider
\begin{align}
    y_{\mu}y^{\mu}+y_5^2=R^2\,,\qquad \mu=1,2,3,4\,,
\end{align}
where
\begin{align}
\label{eq:Yparametrization}
    y^{\mu}=\frac{2R^2x^{\mu}}{(R^2+x^2)}\,,\qquad y^5=\frac{R(R^2-x^2)}{(R^2+x^2)}\,,\qquad   x^2=x_{\mu}x^{\mu}\,.
\end{align}
The conformally flat metric that corresponds to $S^4$ is obtained by the pullback:
\begin{align}\label{eq:EAdSmetric}
    ds^2=\Big(\frac{\pl y^a}{\pl x^{\mu}}\frac{\pl y^b}{\pl x^{\nu}}\eta_{ab}\Big)dx^{\mu}dx^{\nu}:=g_{\mu\nu}dx^{\mu}dx^{\nu}=\frac{4R^4\eta_{\mu\nu}dx^{\mu}dx^{\nu}}{(R^2+x^2)^2} \,,
\end{align}
where $\eta_{\mu\nu}=\diag(+,+,+,+)$. Though this coordinate system does not give us the desired Lorentzian signature as in \cite{Nagaraj:2019zmk}, the metric \eqref{eq:EAdSmetric} admits a smooth flat limit when $R\rightarrow \infty$. 

\paragraph{Almost-commutative twistor space.} Since, $\mso(6)\simeq \msu(4)$ we can consider the maps:
\begin{align}
    Y^{AB}=-Y^{BA}=\ell_p^{-1}Y^a(\gamma_a)^{AB}\,,\qquad L^{AB}= L^{BA}=\frac{1}{2} M^{ab}\Sigma^{AB}_{ab}\, 
     \,,
\end{align}
where 
\begin{align}
    \Sigma_{ab}^{AB}=-\Sigma_{ba}^{AB}=\Sigma_{ab}^{AB}=\frac{i}{4}[\gamma_a,\gamma_b]^{AB}\,
\end{align}
provide the spinorial representation of $\mso(5)\simeq \msp(4)$. The $\msu(4)$ algebra reads \cite{Sharapov:2019pdu}
\begin{subequations}
\begin{align}
    [L^{AB},L^{CD}]&=i(L^{AC}C^{BD}+L^{AD}C^{BC}+L^{BD}C^{AC}+L^{BC}C^{AD})\,,\\
    [L^{AB},Y^{CD}]&=i(Y^{AC}C^{BD}+Y^{BC}C^{AD}-Y^{AD}C^{BC}-Y^{BD}C^{AC})\,,\\
    [Y^{AB},Y^{CD}]&=i(L^{AC}C^{BD}-L^{AD}C^{BC}-L^{BC}C^{AD}+L^{BD}C^{AC})\,.
\end{align}
\end{subequations}
Here, we recognize the $L^{AB}$ as
$\msp(4)$ generators, and $Y^{AB}$ as ``vectors'' that transform  under $\msp(4)$. 
This realization allows us to view $S_N^4$ as a non-commutative twistor space spanned by $\msp(4)$ vectors $Z^A$ and their dual $\hat{Z}^A$. Note that \cite{Penrose:1972ia,Claus:1999xr,Hannabuss:2001xj,Heckman:2011qt}
\begin{align}\label{eq:quantizedCP3}
    i\{Z^A,\hat{Z}^B\}=C^{AB}\,,\qquad \qquad  C^{AB}=\text{diag}(\eps^{\alpha\beta}\,,\eps^{\dot\alpha\dot\beta})\,.
\end{align}
In terms of Weyl spinors, the above relations decompose into
\begin{align}\label{poissonspinors}
    i\{\lambda^{\alpha},\hat\lambda^{\beta}\}=\eps^{\alpha\beta}\,,\qquad \qquad i\{\mu^{\dot\alpha},\hat{\mu}^{\dot\beta}\}=\eps^{\dot\alpha\dot\beta}\,.
\end{align}
It is crucial to note that all Weyl spinors are \underline{dimensionless} in the approach of almost-commutative twistor construction. In addition, the incident relations \eqref{eq:increl} remain the same, i.e.
\begin{align}\label{eq:increldimless}
    \mu^{\dot\alpha}=\tx^{\alpha\dot\alpha}\lambda_{\alpha}\,,
\end{align}
where $\tx^{\alpha\dot\alpha}$ is also dimensionless. Its relation to the dimensionful $x^{\mu}$ reads
\begin{align}\label{eq:xtranslation}
    x^2=\frac{R^2}{4}\tx^2\,,\qquad  \qquad \tx^2:= \tx^{\alpha\dot\alpha}\tx_{\alpha\dot\alpha}\,.
\end{align}

Consider the following symplectic form on almost-commutative twistor space $\P^3_N$:
\begin{align}
    \Omega=d\hat{Z}^A\wedge dZ_{A}=(1+\tx^2)\Big[D\hat{\lambda}^{\alpha}\wedge D\lambda_{\alpha}+\hat{\lambda}_{\alpha}\frac{d\tx^{\alpha\dot\alpha}\wedge d\tx^{\beta}_{\ \dot\alpha}}{(1+\tx^2)^2}\lambda_{\beta}\Big]\,,
\end{align}
where we have used the incident relation \eqref{eq:increl} and
\begin{align}
    D\hat{\lambda}^{\alpha}=d\hat{\lambda}^{\alpha}+\frac{d\tx_{\beta\dot\beta}\tx^{\alpha\dot\beta}}{(1+\tx^2)}\hat{\lambda}^{\beta}\,,\qquad \qquad 
    D\lambda_{\alpha}=d\lambda_{\alpha}+\frac{\tx_{\alpha\dot\beta}d\tx^{\beta\dot\beta}}{(1+\tx^2)}\lambda_{\beta}\,.
\end{align}
This means that $\P^3$ can be identified as the total space of $\P^1$-bundle over $S^4$, which is in the same spirit with \eqref{eq:PT}. Thus, any generating function $f(\tx|Z,\hat{Z})$ on almost commutative twistor space can be written as\footnote{Recall that there will always be an equal number of $\lambda$ and $\hat\lambda$ by virtue of \eqref{eq:spectrumtruncated}.}
\begin{align}
    f(\tx|Z,\hat Z)\rightarrow f(\tx|\lambda,\hat\lambda)=\sum_{i,j}f_{\alpha(i)\beta(i)}(\tx)\lambda^{\alpha(i)}\hat{\lambda}^{\beta(i)}\,.
\end{align}
It is clear from the above that the fuzzy Riemann sphere with coordinates $(\lambda,\hat\lambda)$ is responsible for the internal quantized structure of spacetime. 

The correspondence between almost-commutative twistor space and spacetime is expressed via the Hopf fibration \cite{Sperling:2017gmy}
\begin{align}\label{Hopf-map}
    \begin{split}
    \P^1\xhookrightarrow{}\P^3 \simeq S^7/_{U(1)}&\rightarrow S^4\,,  \\
     Z^{A} &\mapsto  y^{a} :=-\frac {\ell_p}{2}{\hat Z}^{A} (\gamma^{a})_{AB} Z^{B}\,,
   \end{split}
\end{align}
where $S^7$ is defined by \eqref{eq:S7}, and a convenient basis for the $\gamma$-matrices is:
\begin{align}
 (\gamma_{m})^{A}_{\ B}= i
\begin{pmatrix}
 0 & -(\sigma_m)^{\alpha}_{\ \dot\beta} \\ (\sigma_m)_{\beta}^{\ \dot\alpha} & 0
\end{pmatrix}
\qquad
(\gamma_4)_{\ B}^{A} =
\begin{pmatrix}
 0 & \one_2 \\ \one_2 & 0
\end{pmatrix}, 
\qquad
(\gamma_{5})^{A}_{\ B}=
\begin{pmatrix}
 \one_2 & 0 \\ 0 &-\one_2
\end{pmatrix}
\label{gamma-so5-explicit}
\end{align}
for $m=1,2,3$,
where $\sigma_{\mu}=(i\sigma_m,\one_2)$. Another way to express the above basis is to lower their indices down:
 \begin{align}\label{so(5)basisdown}
  (\gamma_{m})_{AB}= 
\begin{pmatrix}
 0 & (\tilde\sigma_m)_{\alpha\dot\beta} \\ 
 -(\tilde\sigma_m)_{\dot\beta\alpha} & 0
\end{pmatrix}\,,
\quad
 (\gamma_{4})_{AB}= 
\begin{pmatrix}
 0 & -\epsilon_{\alpha\dot\alpha} \\ 
 \epsilon_{\alpha\dot\alpha} & 0
\end{pmatrix}\,,
\quad
(\gamma_{5})_{AB}=
\begin{pmatrix}
 -\epsilon_{\alpha\beta} & 0 \\ 0 &\epsilon_{\dot\alpha\dot\beta}
\end{pmatrix}
 \end{align}
where $\tilde\sigma^m_{\alpha\dot\alpha}=-i(\sigma_m)^{\bullet}_{\ \dot\alpha}\epsilon_{\bullet\alpha}$. 
\paragraph{Spinorial representation of the IKKT model.} The twistor representation of the action \eqref{eq:IKKTaction} in the semi-classical limit reads:
\begin{equation}
    \begin{split}
    S=\int \mho\, &\Big(i\{y^{AB},y^{CD}\}\{y_{AB},y_{CD}\}+\bar{\Psi}^A\{Y_{AB},\Psi^B\}\Big)\\
    &\Big(i\{\phi^{\cI\cJ},\phi^{\cM\cN}\}\{\phi_{\cI\cJ},\phi_{\cM\cN}\}+\bar{\Psi}^{\cI}\{\phi_{\cI \cJ},\Psi^{\cJ}\}\Big)\,,
    \end{split}
\end{equation}
where the remaining five coordinates of $SO(10)$ are treated as scalar fields $\phi^{\cI\cJ}$, which carry $\cI,\cJ=1,2,3,4$ indices of the internal symmetry group $SU(4)$ of $\cN=4$ SYM. Since $y_5$ transforms under the external $SO(5)$, the model breaks supersymmetries explicitly when it is placed on non-commutative twistor space. However, note that the internal group $SU(4)$ can be recovered in the flat limit, see e.g. \cite{Steinacker:2010rh}. 

From \eqref{so(5)basisdown}, we can decompose
\begin{align}\label{eq:YABdecomposition}
    y^{AB}=p^{AB}+q^{AB}=\begin{pmatrix}
     0 & p^{\alpha\dot\beta}\\
     -p^{\dot\beta\alpha}&0
    \end{pmatrix}+\begin{pmatrix}
     q^{\alpha\beta} & 0\\
     0& q^{\dot\alpha\dot\beta}
    \end{pmatrix}\,,
\end{align}
where $p^{AB}$ are off-diagonal, and $q^{AB}$ stands for the fifth direction. In particular,
\begin{subequations}
\begin{align}
    p^{\alpha\dot\alpha}&=p^{\mu}\hat{\sigma}_{\mu}^{\alpha\dot\alpha}\,,\qquad \qquad  \hat\sigma_{\mu}^{\alpha\dot\alpha}=(i\sigma_3,\one{_2},-i\sigma_1,i\sigma_2)\\
    q^{\alpha\beta} &= - y_5\epsilon^{\alpha\beta} , \qquad \qquad \ \,
     q^{\dot\alpha\dot\beta} = + y_5\epsilon^{\dot\alpha\dot\beta} \,.
\end{align}
\end{subequations}

Now, we can consider the following decompositions of $p$ and $q$ in terms of `background' plus fluctuations.
\begin{align}
    \binom{p^{\alpha\dot\alpha}}{q^{\alpha\beta}}=\binom{\ty^{\alpha\dot\alpha}}{\ty_5\epsilon^{\alpha\beta}}+\binom{A^{\alpha\dot\alpha}}{\hat\phi\epsilon^{\alpha\beta}}\,,
\end{align}
where $\hat\phi$ is the scalar field that the external $SO(5)$ acts on. 

For simplicity, let us consider the semi-classical and flat limit of the IKKT matrix. It can be checked that all contributions related to the background $\ty_5$ vanish in this limit \cite{Steinacker:2022jjv}. Furthermore, the external scalar $\hatphi$ will rejoin with the other 5 internal scalars to form the 6 adjoint scalars of the internal group $SU(4)$. Due to the non-commutativity of coordinates/fields, there will be more terms (some of which we do not have a clear interpretation of) in the IKKT action, as illustrated in \cite{Sperling:2017dts,Sperling:2017gmy}. This makes it difficult to deal explicitly with the mixtures between backgrounds and fluctuations. For illustrative reason, consider the `Yang-Mills' term in the IKKT matrix model where
\begin{equation}
    \begin{split}
    \frac{1}{2}F_{\alpha\alpha}F^{\alpha\alpha}&=2\{\ty^{\alpha}{}_{\dot\gamma},A^{\alpha\dot\gamma}\}\{\ty_{\alpha\dot\zeta},A_{\alpha}{}^{\dot\zeta}\}+\{\ty^{\alpha}{}_{\dot\gamma},\ty^{\alpha\dot\gamma}\}\{A_{\alpha\dot\zeta},A_{\alpha}{}^{\dot\zeta}\}\\
    &+2\{\ty^{\alpha}{}_{\dot\gamma},A^{\alpha\dot\gamma}\}\{A_{\alpha\dot\zeta},A_{\alpha}{}^{\dot\zeta}\}+\frac{1}{2}\{A^{\alpha}{}_{\dot\gamma},A^{\alpha\dot\gamma}\}\{A_{\alpha\dot\zeta},A_{\alpha}{}{}^{\dot\zeta}\}\,.
\end{split}
\end{equation}
The troublesome term that we mentioned about is the second term in the above expression. To circumvent this problem, one can introduce an auxiliary $B_{\alpha\alpha}$ field to absorb these troublesome terms:
\begin{align}
    S^{\text{YM}}=\int \mho\, \Big(B_{\alpha\alpha}F^{\alpha\alpha}+\frac{i}{2}B_{\alpha\alpha}B^{\alpha\alpha}\Big)\,.
\end{align}
As a result, we obtain the following action for the IKKT-matrix model in the semi-classical and flat limit
\begin{equation}\label{fullIKKT}
    \begin{split}
    S=\int \mho \Big[&B_{\alpha\alpha}F^{\alpha\alpha}+\frac{i}{2}B_{\alpha\alpha}B^{\alpha\alpha}+i\{p^{\alpha\dot\alpha},\phi^{\cI\cJ}\}\{p_{\alpha\dot\alpha},\phi_{\cI\cJ}\}+2\bar{\chi}^{\alpha}\{p_{\alpha\dot\beta},\widetilde\chi^{\dot\beta}\}\\
    &+\bar{\chi}^{\cI}\{\phi_{\cI\cJ},\chi^{\cJ}\}+\widetilde{\bar{\chi}}^{\cI}\{\phi_{\cI\cJ},\widetilde\chi^{\cJ}\}+\frac{i}{2}\{\phi^{\cI\cJ},\phi^{\cM\cN}\}\{\phi_{\cI\cJ},\phi_{\cM\cN}\}\Big]\,,
    \end{split}
\end{equation}
where $\Psi^A=(\chi^{\alpha},\widetilde\chi^{\dot\alpha})$ and $\Psi^{\cI}=(\chi^{\cI},\widetilde\chi^{\cI})$. Note that the spacetime action of the IKKT model
is already recognized in the above action without the need of referring to the twistor cohomology. This is main advantage of this approach compared to the previous section. 

Note that by dropping some terms in \eqref{fullIKKT} while retaining gauge-invariance, we can obtain the self-dual sector of IKKT-matrix model:
\begin{equation}\label{selfdualIKKT}
    \begin{split}
    S_{SD}=\int \mho \Big[&B_{\alpha\alpha}F^{\alpha\alpha}+i\{p^{\alpha\dot\alpha},\phi^{\cI\cJ}\}\{p_{\alpha\dot\alpha},\phi_{\cI\cJ}\}+2\bar{\chi}^{\alpha}\{p_{\alpha\dot\beta},\widetilde\chi^{\dot\beta}\}+\widetilde{\bar{\chi}}^{\cI}\{\phi_{\cI\cJ},\widetilde\chi^{\cJ}\}\Big]\,.
    \end{split}
\end{equation}

To obtain the spacetime action of the twistor actions \eqref{fullIKKT} and \eqref{selfdualIKKT}, one can proceed as before by integrating out all fiber coordinates. However, before doing that let us show the structure of interactions coming from the Poisson brackets $\{\,,\}$.

\paragraph{Spinorial effective vielbein and derivativation.} Essentially, since everything can be expressed in terms of spinors, all derivatives thereof can be obtained by \eqref{poissonspinors}. For example, if we consider $\ty^{\alpha\dot\alpha}=-(\hat{\lambda}^{\alpha}\mu^{\dot\alpha}-\lambda^{\alpha}\hat{\mu}^{\dot\alpha})$, then
\begin{subequations}\label{eq:Poissononcoordinates2}
\begin{align}
     \{\ty^{\alpha\dot\alpha},\lambda^{\beta}\}
    &=+i\epsilon^{\alpha\beta}\mu^{\dot\alpha}\,,\\
    \{\ty^{\alpha\dot\alpha},\hat{\lambda}^{\beta}\}&=-i\epsilon^{\alpha\beta}\hat{\mu}^{\dot\alpha}\,.
\end{align}
\end{subequations}
As a result,
\begin{align}\label{eq:Poissononcoordinates1}
    \begin{split}
    \{\ty^{\alpha\dot\alpha},\ty^{\beta\dot\beta}\}&= 2i(\lambda^{(\alpha}\hat{\lambda}^{\beta)}\epsilon^{\dot\alpha\dot\beta}+\mu^{(\dot\alpha}\hat{\mu}^{\dot\beta)}\epsilon^{\alpha\beta})\,,
     \end{split}
\end{align}
Using the above information, we can define
\begin{align}\label{nabla-def-1}
    \begin{split}
    \{\ty^{\alpha\dot\alpha},\varphi(\ty|\lambda,\hat\lambda)\}:&=\Big(\{\ty^{\alpha\dot\alpha},\ty^{\beta\dot\beta}\}\frac{\pl}{\pl \ty^{\beta\dot\beta}}+\{\ty^{\alpha\dot\alpha},\lambda^{\beta}\}\frac{\pl}{\pl \lambda^{\beta}}+\{\ty^{\alpha\dot\alpha},\hat{\lambda}^{\beta}\}\frac{\pl}{\pl \hat{\lambda}^{\beta}}\Big)\varphi\,\\
    &=\cE^{\alpha\dot\alpha|\beta\dot\beta}\pl_{\beta\dot\beta}\varphi+\cE^{\alpha\dot\alpha|\beta}\frac{\pl}{\pl \lambda^{\beta}}\varphi+\hat{\cE}^{\alpha\dot\alpha|\beta}\frac{\pl}{\pl \hat{\lambda}^{\beta}}\varphi\,.
    \end{split}
\end{align}
The objects $\cE$ are referred to as the effective spinorial vielbeins. There are two $\cE$ that are particularly important to us since they are used to construct the effect metric:
\begin{subequations}\label{eq:effectivevielbein}
\begin{align}
    \cE^{\alpha\dot\alpha|\beta\dot\beta}:&=\{\ty^{\alpha\dot\alpha},\ty^{\beta\dot\beta}\}=2i(\lambda^{(\alpha}\hat{\lambda}^{\beta)}\epsilon^{\dot\alpha\dot\beta}+\mu^{(\dot\alpha}\hat{\mu}^{\dot\beta)}\epsilon^{\alpha\beta})\,\label{eq:1steffectivevielbein},\\
    \cE^{5|\alpha\dot\alpha}:&=\{\ty^5,\ty^{\alpha\dot\alpha}\}=-i(\hat{\lambda}^{\alpha}\mu^{\dot\alpha}+\lambda^{\alpha}\hat{\mu}^{\dot\alpha})\,.
\end{align}
\end{subequations}
Note that in the flat limit, all expression involved dotted spinor $\mu,\hat\mu$ will vanish (recall that the incident relations \eqref{eq:increldimless} contain dimensionless spinors; so $\mu$ roughly scales as $1/\sqrt{R}$ in this picture).
\paragraph{The effective metric.} Consider some scalar field $\vartheta(\ty)$ and the kinetic term 
\begin{align}\label{eq:preeffectivemetric}
    \begin{split}
    \{\ty^{\zeta\dot\zeta},\vartheta\}\{\ty_{\zeta\dot\zeta},\vartheta\}+\{\ty^5,\vartheta\}\{\ty^5,\vartheta\}&=\cE^{\zeta\dot\zeta|\alpha\dot\alpha}\pl_{\alpha\dot\alpha}\vartheta\, \cE_{\zeta\dot\zeta|\beta\dot\beta}\pl^{\beta\dot\beta}\vartheta+\cE^{5|\alpha\dot\alpha}\pl_{\alpha\dot\alpha}\vartheta\,\cE_{5|\beta\dot\beta}\pl^{\beta\dot\beta}\vartheta\\
    &=:G^{\alpha\dot\alpha\beta\dot\beta}\pl_{\alpha\dot\alpha}\vartheta \,\pl_{\beta\dot\beta}\vartheta\,.
    \end{split}
\end{align}
This gives us the effective metric:
\begin{align}\label{fullmetric}
    G^{\alpha\dot\alpha\beta\dot\beta}(\ty)=N^2\epsilon^{\alpha\beta}\epsilon^{\dot\alpha\dot\beta}-\ty^{\alpha\dot\alpha}\ty^{\beta\dot\beta}\,.
\end{align}
In terms of $\tx^{\alpha\dot\alpha}\in S^4$, the effective metric reads
\begin{align}\label{eq:effectivemetricx}
    G^{\alpha\dot\alpha\beta\dot\beta}(\tx)=\langle \lambda\hat\lambda \rangle^2\Big(\frac{N^2}{\langle \lambda \hat\lambda\rangle^2}\epsilon^{\alpha\beta}\epsilon^{\dot\alpha\dot\beta}-\tx^{\alpha\dot\alpha}\tx^{\beta\dot\beta}\Big)\,.
\end{align}

\paragraph{EoM, DoF and plane-wave solution in the flat limit.} Since we started with a 5-dimensional system, it is important to ask whether the higher-spin spacetime fields will carry only two degrees of freedom. To answer this question, it is good enough to look at the free action in \eqref{selfdualIKKT} and extract the free equations of motion for the higher-spin valued gauge fields $A^{\alpha\dot\alpha}$ and $B_{\alpha\alpha}$:
\begin{subequations}
\begin{align}
    \{\ty^{\alpha}{}_{ \dot\alpha},A^{\alpha\dot\alpha}\}&=0\,, \qquad \delta A^{\alpha\dot\alpha}=\{\ty^{\alpha\dot\alpha},\xi\}\,,\\
    \{\ty^{\gamma}{}_{ \dot\alpha},B_{\gamma\alpha}\}&=0\,.
\end{align}
\end{subequations}
Here, $\xi$ is some $\hs$-valued section on $\Kcurl=
(\P^1)^{N}\times \R^4$. As alluded to above, $A^{\alpha\dot\alpha}$ is a $\hs$-valued \textit{function}; and it has the following mode expansion
\begin{align}
A^{\alpha\dot\alpha}(\tx)= \sum_s A^{\kappa(s)\tau(s)|\alpha\dot\alpha}(\tx)\lambda_{\kappa(s)}\hat{\lambda}_{\tau(s)}\,,
\label{planewave-gaugemodes-C}
\end{align}
with $(\alpha,\dot\alpha)$ are independent indices. This gives 4 independent tangential modes of the $A$ field \cite{Sperling:2018xrm}. We can decompose $A^{\alpha\dot\alpha}$ into two eigenmodes
\begin{subequations}
\begin{align}
A_{(1)}^{\alpha\dot\alpha} \ &= 
A^{(\kappa(2s)\alpha)\dot\alpha}\lambda_{\kappa(s)}\hat{\lambda}_{\kappa(s)}\,, \\
A_{(2)}^{\alpha\dot\alpha} \ &= \ \epsilon^{\alpha\kappa} \TA^{\kappa(2s-1)\dot\alpha}\lambda_{\kappa(s)}\hat{\lambda}_{\kappa(s)}=\lambda^{\alpha}\TA^{\kappa(2s-1)\dot\alpha}\lambda_{\kappa(s-1)}\hat{\lambda}_{\kappa(s)}+\hat{\lambda}^{\alpha}\TA^{\kappa(2s-1)\dot\alpha}\lambda_{\kappa(s)}\hat{\lambda}_{\kappa(s-1)} \,.
\end{align}
\label{A-ansatz-1-2}
\end{subequations}
The gauge field $A^{\alpha\dot\alpha}$ transforms as
\begin{align}
    \delta_{\xi,\vartheta}A^{\kappa(2s)|\alpha\dot\alpha}=\{\ty^{\alpha\dot\alpha},\xi^{\kappa(2s)}\}+\epsilon^{\kappa\alpha}\vartheta^{\kappa(2s-1)\dot\alpha}\,.
\end{align}
where $\xi$ and $\vartheta$ are some $\hs$-valued sections on $\Kcurl$. Here, the algebraic symmetry, whose gauge parameter is $\vartheta$, is used to gauge away the (unwanted) second eigenmode $A_{(2)}$.

On the other hand, the $B_{\alpha\alpha}$ field can be decomposed as
\begin{subequations}
\begin{align}
    B_{(1)}^{\alpha\bullet}&=B^{\kappa(2s)\alpha\bullet}\lambda_{\kappa(s)}\hat\lambda_{\kappa(s)}\,,\label{B-mode-1}\\
    B_{(2)}^{\alpha \bullet}&=\epsilon^{\bullet \kappa}\TB^{\kappa(2s-1)\alpha}\lambda_{\kappa(s)}\hat\lambda_{\kappa(s)}\,, \label{B-mode-2}
\end{align}
\end{subequations}
where $\bullet$ is the index that contract to $\ty^{\bullet}{}_{\dot\alpha}$. Note that the second mode $B_{(2)}^{\alpha\alpha}$ plays the role of a Lagrangian multiplier and provides us a non-commutative version of the Lorenz gauge condition:
\begin{align}
    \int \mho \,\TB_{\alpha(2n-1)}\{\ty_{\alpha\dot\alpha},A^{\alpha(2n)
    \dot\alpha}\}\,.
\end{align}
Therefore, only the first eigenmodes of $A^{\alpha\dot\alpha}$ and $B_{\alpha\alpha}$ propagate. Since there are $2s+1$ equations in $\{\ty^{\alpha}{}_{ \dot\alpha},A^{\alpha(2s-1)\dot\alpha}\}$ and there are $2s-1$ number of components in the gauge symmetry generator $\xi$, the number of degree of freedom the $A_{(1)}^{\alpha\dot\alpha}$ eigenmodes has is \cite{Kaparulin:2012px}:
\begin{align}
    \frac{2s+1-(2s-1)}{2}=1\,.
\end{align}
For the $B$ field, there are in total $4s$ equations in $\{\ty^{\gamma}{}_{\dot\alpha},B_{\gamma\alpha(2s-1)}\}$ and there are $(2s-1)$ 2nd order fuzzy Bianchi identities
\begin{align}
    \{\ty^{\gamma\dot\alpha},\{\ty^{\gamma}{}_{ \dot\alpha},B_{\gamma\gamma\alpha(2s-2)}\}\}\simeq 0\,.
\end{align}
Thus, the number of degree of freedom that $B_{(1)}$ possesses is
\begin{align}
    \frac{4s-2(2s-1)}{2}=1\,.
\end{align}
Together, $A$ and $B$ (which correspond to positive/negative helicity `spacetime' fields) describe massless higher-spin fields in complexified Euclidean spacetime. Note that besides the affine patch considered in \cite{Steinacker:2022jjv}, there is also the FLRW patch where the intrinsic signature is Minkowski \cite{Sperling:2019xar,Steinacker:2019awe}. This is the patch that is relevant for matrix-model type cosmology.

Next, we can use \eqref{poissonspinors} to solve the free equations of motion for the $A^{\alpha\dot\alpha}$ and $B_{\alpha\alpha}$ fields. First of all,
\begin{equation}\label{eom-A1-mode}
    \begin{split}
    0=\{\ty^{\alpha}{}_{ \dot\alpha},A_{(1)}^{\alpha\dot\alpha}\} 
    &=  \{\ty^{\alpha}{}_{\dot\alpha}, A^{\kappa(2s)\alpha\dot\alpha}(\tx)\lambda_{\kappa(s)}\hat{\lambda}_{\kappa(s)}\} \\
    &=\frac{1}{\langle \lambda \hat\lambda\rangle} \cE_{\ \, \dot\alpha}^{\alpha \  |\beta\dot\beta}\frac{\partial}{\partial \tx^{\beta\dot\beta}}A^{\kappa(2s)\alpha\dot\alpha}(\tx)
    \lambda_{\kappa(s)}\hat{\lambda}_{\kappa(s)}\,.
    \end{split}
\end{equation}
This is equivalent to saying that
\begin{align}
    \cE_{\ \, \dot\alpha}^{\alpha \  |\beta\dot\beta}\frac{\partial A^{\kappa(2s)\alpha\dot\alpha}}{\pl \tx^{\beta\dot\beta}} 
    \simeq
    \lambda^{(\alpha}\hat{\lambda}^{\beta)}\frac{\partial A^{\kappa(2s)\alpha\dot\alpha}}{\partial \tx^{\beta\dot\alpha}}
    =0
\end{align}
in the flat limit. The solution of the above reads
\begin{align}\label{eq:positivehelicityansatz}
    A^{\alpha(2s+1)\dot\alpha}=\frac{\zeta^{\alpha}...\zeta^{\alpha}\tilde{\upsilon}^{\dot\alpha}}{\langle \zeta \upsilon\rangle^{2s+1}}\,e^{i\upsilon^{\alpha}\tx_{\alpha\dot\alpha}\tilde{\upsilon}^{\dot\alpha}}\,
\end{align}
in terms of on-shell four momentum $k^{\alpha\dot\alpha}=\upsilon^{\alpha}\tilde\upsilon^{\dot\alpha}$.\footnote{We assume there is a suitable analytic continuation or complexification.} Note that $A_{(1)}$ satisfies the gauge-fixing condition 
\begin{align}
     \frac{\pl}{\pl \tx^{\alpha\dot\alpha}}A_{(1)}^{\alpha\dot\alpha} = 0\,.
     \label{A1-solutions-gaugefix}
\end{align}
For the $B_{(1)}^{\alpha\alpha}$ eigenmodes, we find
\begin{align}\label{eq:negativehelicityansatz}
    B^{\alpha(2s)}=\upsilon^{\alpha}...\upsilon^{\alpha}e^{i\upsilon^{\kappa}\tx_{\kappa\dot\kappa}\tilde{\upsilon}^{\dot\kappa}}\,.
\end{align}

\paragraph{Reduction to spacetime.} Let us now show the structure of the cubic vertices $B\{A,A\}$ for the self-dual YM sector of the HS-IKKT in spacetime. Since $\{A^{\alpha}{}_{ \dot\alpha},A^{\alpha\dot\alpha}\}$  involves both the fibers and spacetime derivatives on $\Kcurl=(\P_1)^N\times \R^4$, there will be more terms in spacetime compared to the holomorphic case (cf., \eqref{eq:holomorphicPoisson}) due to the structure of the effective vielbein $\cE$. However, as discussed, the advantage of this almost-commutative twistor construction is that everything is naturally higher-spin extensible and the fibers does not get deformed by the Poisson bracket \eqref{eq:quantizedCP3}. The leading contributions in $B\{A,A\}$ read
\begin{align}
    \begin{split}
    &\lambda^{\beta(s)}\hat{\lambda}^{\rho(s)}\,B_{\alpha\alpha\beta(s)\rho(s)}\lambda_{\gamma(m)}\hat{\lambda}_{\delta(m)}\lambda_{\kappa(n)}\hat{\lambda}_{\tau(n)}\{A^{\gamma(m)\delta(m)\alpha}{}_{\dot\alpha},A^{\kappa(n)\tau(n)\alpha\dot\alpha}(\tx)\}\\
    &\sim\lambda^{\beta(s)}\hat{\lambda}^{\rho(s)}B_{\alpha\alpha\beta(s)\rho(s)}\lambda_{\gamma(m)}\hat{\lambda}_{\delta(m)}\lambda_{\kappa(n)}\hat{\lambda}_{\tau(n)}\cE^{\circ\dot\circ\bullet\dot\bullet}\pl_{\circ\dot\circ}A^{\gamma(m)\delta(m)\alpha}{}_{\dot\alpha}\,\pl _{\bullet\dot\bullet}A^{\kappa(n)\tau(n)\alpha\dot\alpha}\,.
    \end{split}
\end{align}
Using \eqref{eq:bridge}, we obtain the following spacetime expression
\begin{align}\label{eq:thevertices}
   V_3^{\text{lead}}=\sum_{m+n=2s-2} B_{\alpha(2s)}\pl_{\alpha \dot\bullet}A^{\alpha(m)}{}_{\dot\alpha}\pl_{\alpha}{}^{\dot\bullet}A^{\alpha(n)\dot\alpha}
   +V^{\text{irrelevant}}\,.
\end{align}
Here, $V^{\text{irrelevant}}$ are contributions that vanish when we plugging in the asymptotic states \eqref{eq:positivehelicityansatz} and \eqref{eq:negativehelicityansatz}. It is remarkable that the cubic vertex \eqref{eq:thevertices} is closely related to the one of the higher-spin extension of self-dual gravity in \cite{Krasnov:2021nsq}. For this reason, HS-IKKT is a gravitational  theory of higher spins.

Besides the leading term, we also get the following subleading contributions:
\begin{align}\label{eq:V3sub}
    V_3^{\text{sub}}=2\sum_{m+n=2s-2}mn\,B_{\alpha(2s-2)}A^{\alpha(m)}{}_{ \bullet\dot\bullet}A^{\alpha(n)\bullet\dot\bullet}\,.
\end{align}
Their contributions to the scattering amplitudes vanish upon substituting the plane-wave solutions \eqref{eq:positivehelicityansatz}. Furthermore, one can check that this vertex vanishes identically in the light-cone gauge. As an observation, we see that the only non-vanishing contributions coming from the Poisson bracket between two twistor fields contains terms with derivatives acting on fields but the fiber coordinates.

\section*{Acknowledgement}
It is a pleasure to thank Tim Adamo, Roland Bittleston, Atul Sharma, Zhenya Skvortsov and Harold Steinacker for useful discussions. The author also appreciates the organisers of the Corfu Summer Institute 2022 and the Humboldt Kolleg on “Noncommutative and generalized geometry in string theory, gauge theory and related physical models” the invitation to deliver a talk and stimulating environment for discussions. This work is partially supported by the Fonds de la Recherche Scientifique under Grants No. F.4503.20 (HighSpinSymm), Grant No. 40003607 (HigherSpinGraWave), T.0022.19 (Fundamental issues
in extended gravitational theories) and the funding from the European Research Council (ERC) under Grant No. 101002551.

\setstretch{0.5}
\footnotesize
\bibliography{twistor.bib}
\bibliographystyle{JHEP-2}

\end{document}